\documentclass[reprint,nofootinbib,pra,twocolumn]{revtex4}
\pdfoutput=1
\usepackage{graphicx}
\usepackage{dcolumn}
\usepackage{bm}

\usepackage{amsmath}
\usepackage{amssymb}

\begin{document}

\title{Diffusion, thermalization and optical pumping of YbF molecules in a cold buffer gas cell}

\author{S. M. Skoff}
\author{R. J. Hendricks}
\author{C. D. J. Sinclair}
\author{J. J. Hudson}
\author{D. M. Segal}
\author{B. E. Sauer}
\author{E. A. Hinds}
\author{M. R. Tarbutt}\email{m.tarbutt@imperial.ac.uk}

\affiliation{Centre for Cold Matter, Blackett Laboratory, Imperial College London, Prince Consort Road, London, SW7 2AZ. United Kingdom.}

\begin{abstract}
We produce YbF molecules with a density of $10^{18}$\,m$^{-3}$ using laser ablation inside a cryogenically-cooled cell filled with a helium buffer gas. Using absorption imaging and absorption spectroscopy we study the formation, diffusion, thermalization and optical pumping of the molecules. The absorption images show an initial rapid expansion of molecules away from the ablation target followed by a much slower diffusion to the cell walls. We study how the time constant for diffusion depends on the helium density and temperature, and obtain values for the YbF-He diffusion cross-section at two different temperatures. We measure the translational and rotational temperatures of the molecules as a function of time since formation, obtain the characteristic time constant for the molecules to thermalize with the cell walls, and elucidate the process responsible for limiting this thermalization rate. Finally, we make a detailed study of how the absorption of the probe laser saturates as its intensity increases, showing that the saturation intensity is proportional to the helium density. We use this to estimate collision rates and the density of molecules in the cell.
\end{abstract}
\maketitle


\section{Introduction}
\label{sec:introduction}

Cold molecules are useful for a wide variety of applications in physics and chemistry. A review of these applications and the current status of the field is given in reference \cite{Carr:2009}. An important application is the use of cold molecules for testing fundamental laws of physics \cite{Tarbutt(1):2009}. Cold molecules are being used to test the standard model of particle physics and its extensions via a measurement of the electron's electric dipole moment \cite{Hudson:2002, Tarbutt(2):2009, Vutha:2010}. They can also be used to test the time invariance of the fine structure constant and of the proton-to-electron mass ratio \cite{Hudson:2006, vanVeldhoven:2004, Schiller:2005, Bethlem:2008, Bethlem:2009}, test Lorentz invariance \cite{Muller:2004}, and measure parity-violating interactions in nuclei \cite{DeMille:2008} and in chiral molecules \cite{Darquie:2010}.

Buffer gas cooling is a very versatile way of producing high densities of atoms and molecules at temperatures down to a few hundred milliKelvin \cite{Campbell:2009}. The molecules are introduced into a cryogenically-cooled cell where they thermalize with a cold buffer gas. The buffer gas is usually helium, though neon has also been used \cite{Patterson:2009}. The molecules can be brought into the cell in a number of ways, depending on their vapour pressure and reactivity. Methods that have been demonstrated include laser ablation of a precursor target e.g. \cite{Weinstein(1):1998}, injection through a capillary e.g. \cite{Messer:1984}, loading from a molecular beam e.g. \cite{Egorov:2004}, and direct gas flow loading e.g. \cite{Patterson:2009}.

Once loaded, there are three main options available. In the simplest case the molecules are allowed to diffuse freely to the cell walls, where they will stick, and measurements are made on the molecules in the interval of time between loading and sticking. This method has been used to measure molecular parameters by millimetre and submillimetre spectroscopic techniques \cite{Messer:1984, Willey(1):1988, Willey(2):1988}, and more recently by laser spectroscopy, e.g. \cite{Egorov:2001, Skoff:2009}. Elastic and inelastic collision cross-sections can also be measured this way, e.g. \cite{Lu:2008, Lu:2009}. A second approach is to trap the molecules magnetically by placing the cell inside a strong magnetic trapping field and pumping away the helium as soon as the molecules are cold \cite{Doyle:1995}. The coldest molecules are then confined in the magnetic trap. Several molecular species have been trapped this way, \cite{Weinstein(2):1998, Stoll:2008}, allowing elastic collision and Zeeman relaxation rates to be measured at low temperatures, \cite{Maussang:2005}. Trapping times of 20\,s have been demonstrated for NH molecules \cite{Tsikata:2010}. A third approach is to extract the molecules from the cell to make a high intensity cold beam \cite{Maxwell:2005} which could then be used in a molecular beam experiment or for trapping. The molecules that form the beam can be guided away from the cell and separated from the helium using a magnetic guide \cite{Patterson:2007}, an electrostatic guide \cite{vanBuuren:2009} or an alternating gradient electric guide \cite{Wall:2009}.

It is important to have a good understanding of the processes that occur inside the buffer gas cell, since this is the starting point for all experiments whether they be done directly inside the cell, in a magnetic trap loaded from the cell, or in a beam extracted from the cell. In this paper we focus on the formation, diffusion and thermalization of YbF molecules produced by laser ablation of a target inside a closed helium buffer gas cell. These molecules are of particular interest for the measurement of the electron's electric dipole moment \cite{Hudson:2002, Tarbutt(2):2009} and, for that application, the formation of a high density of YbF is one of our aims. The work presented here will be applicable to many other buffer gas cooling experiments, particularly those where laser ablation is the loading method or where laser spectroscopy is used to make measurements inside the cell.

\section{Experimental Details}
\label{sec:ExperimentSetup}

\begin{figure}
\begin{centering}
\includegraphics[width=0.4\textwidth]{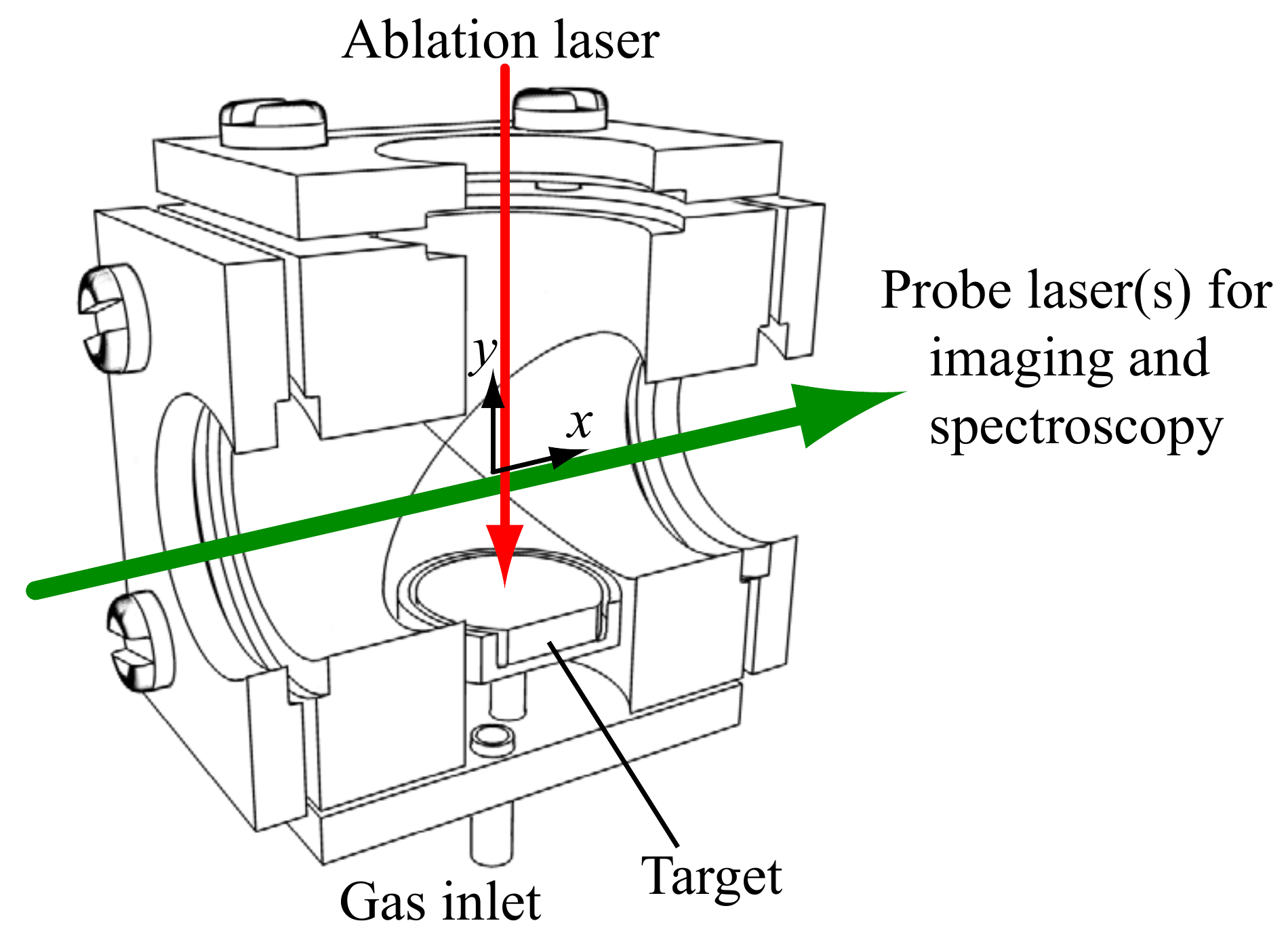}
\caption{\label{fig:ExperimentSetup} (Color online) An illustration of the cell showing the position of the target and the gas inlet and the propagation directions of the ablation and probe lasers. The coordinate system we use is also shown and has its origin at the centre of the cell.}
\end{centering}
\end{figure}

Figure \ref{fig:ExperimentSetup} shows a few details of the experiment. To make the cell we took a solid aluminium cube of side 40\,mm and bored three orthogonal 20\,mm diameter holes, two passing all the way through the cell and the other half way through. The face without a hole mounts onto the cold plate of a liquid helium cryostat (Infrared Laboratories, model HD3). Vacuum-tight indium seals close the five open faces, three with windows and two with blanks. One window is used for laser ablation of a target placed on the opposite side of the cell and the other two are used for laser absorption spectroscopy and imaging of the molecules formed inside. The cryostat is pumped by a turbomolecular pump to a pressure of approximately 10$^{-6}$ mbar. A radiation shield, in thermal contact with a liquid nitrogen dewar, surrounds the setup and prevents room temperature radiation from reaching the cold surfaces. Windows in this shield, coated with indium tin oxide, provide optical access at visible wavelengths, whilst reflecting thermal radiation.

YbF molecules are created in the cell by laser ablation of a vacuum hot-pressed target consisting of 70\% Yb and 30\% AlF$_3$ by mass. The ablation pulses have a duration of 8\,ns, an energy of approximately 50\,mJ, a wavelength of 1064\,nm, and a spot size at the target of 2\,mm.

Helium gas enters the cell through a hole beneath the target, via a thin-walled tube of inner diameter 2.5\,mm. This is in two parts: a stainless steel tube for most of the length is connected to a copper tube near the cold plate of the cryostat. To ensure that the gas is cold when it enters the cell, the copper tubing winds around, and is silver-soldered to, a copper bobbin connected to the cold plate. At the room temperature end, outside the cryostat, the feed tube is connected to a chamber where the helium pressure is measured and controlled. The helium enters this chamber through a leak valve and is pumped away to a rotary pump through a second valve. These two valves provide fine control over the helium pressure, which is measured by a Pirani gauge. When the cell is cold, there is a thermomolecular pressure drop between the two ends of the tube, and we account for this using the empirical formula given in \cite{Roberts:1956}. Even when the cell is at room temperature, it is not immediately clear whether the pressure measured at the inlet end of the feed tube is the same as the pressure in the cell. To investigate this, we temporarily connected a second Pirani gauge directly to the cell, via a port in the cryostat usually used as a window. Both pressure gauges were calibrated to a third gauge which, for helium, is accurate to 0.5\%.

\begin{figure}
\begin{centering}
\includegraphics[width=0.45\textwidth]{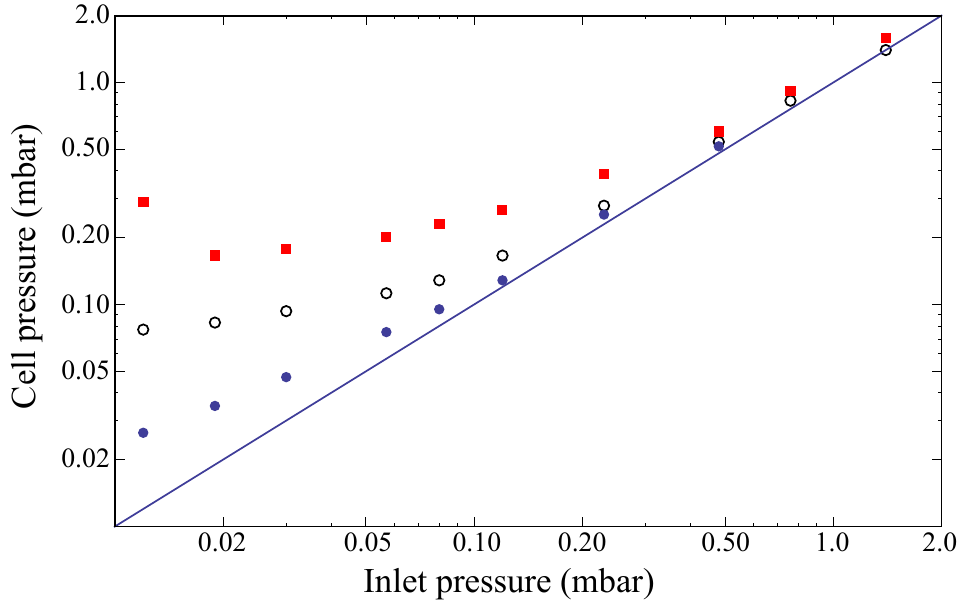}
\caption{\label{fig:pgraph} (Color online) Pressure measured at the cell versus pressure measured at the inlet tube, all at room temperature. Red squares: during ablation. Black open circles: pressure settled after ablation. Blue filled circles: after pumping on the cell through the gas inlet tube for about 12 hours. The errors are about 1\% of the reading. The line has zero offset and a gradient of 1. Note the logarithmic scale.}
\end{centering}
\end{figure}

Figure \ref{fig:pgraph} shows the pressure in the cell versus the pressure at the inlet. One set of measurements was taken during ablation of the target at 10\,Hz, another set a few minutes after stopping the ablation, and a final set after pumping the cell through the gas feed tube for about 12 hours. The figure shows that the cell pressure is equal to the inlet pressure plus a constant offset. The offset is due to outgassing of the target which raises the total pressure in the cell. This outgassing is greatest during ablation of the target, when the offset is $0.17\pm 0.05$\,mbar. When the ablation is stopped the offset falls with two separate timescales. First it falls quickly, reaching a steady value of $0.06 \pm 0.02$\,mbar in a few minutes, and then it falls much more slowly, reaching $0.015 \pm 0.003$\,mbar after about 12 hours. We could only make these direct pressure measurements at room temperature, so we do not know how the offset changes with temperature. The pressure offset has implications for some of our measurements, and we will highlight these where relevant.

The temperature is measured using a silicon diode temperature sensor attached to the cold plate of the cryostat. At the lowest temperatures however, temperature measurements of the fully thermalized molecules (Sec.\,\ref{sec:Thermalisation}) indicate that the cell is about 10\,K hotter than the cold plate, presumably due to insufficient thermal conduction between the two. We have also used the Doppler broadening of the atomic Yb spectrum to determine the temperature, and this agrees with the results obtained from the molecules. Using a new experimental setup, where the thermal conduction is improved, we find that the molecules do indeed reach the same temperature as the 4\,K cell. All the measurements reported here were done with the earlier apparatus and we use the temperature of the molecules as the most reliable measure of the cell temperature.

All the measurements presented in this paper are based on absorption of a cw probe laser beam. Light from a dye laser is used to address individual rotational lines within the (0--0) band of the $X {}^2\mathrm{\Sigma}^+$--$A {}^2\mathrm{\Pi}_{1/2}$ transition of YbF, at a wavelength of 552\,nm. The laser frequency is measured with an accuracy of approximately 600\,MHz using a wavemeter (HighFinesse WS6). Absorption images reveal the spatial distribution of the molecules as a function of time. For these measurements, the laser beam passes through a 100\,MHz acousto-optic modulator (AOM) acting as a fast optical shutter, and is then collimated and expanded to a diameter of 20\,mm. This beam passes through the buffer gas cell and onto a ccd camera (Marlin F033B). The shortest exposure time offered by this camera is 34\,$\mu$s. To take images with higher time resolution, we use the AOM to turn on the absorption imaging beam for a period of 10\,$\mu$s. For time-resolved measurements of the YbF density, translational temperature and rotational temperature at a specific point in the cell, we use a combination of Doppler-limited and Doppler-free absorption spectroscopy, using 1\,mm diameter pump, probe and reference beams, as described in detail in \cite{Skoff:2009}.

Data are taken at a repetition rate of 10Hz, synchronised to the 50Hz line frequency to suppress the effects of line noise in the experiment. In the imaging experiments the laser frequency is fixed and each shot yields one image from the ccd camera. In the spectroscopy experiments, the outputs of the photodiodes are recorded for 8\,ms with a sample rate of 250\,kHz, and the laser frequency is stepped between one shot and the next. As described in \cite{Skoff:2009}, the absorption of a fixed frequency probe beam from a second dye laser is used for normalization purposes, factoring out the slow drift in ablation yield that occurs as the target degrades.

\section{Absorption imaging}
\label{sec:Imaging}

\begin{figure*}
\begin{centering}
\includegraphics[width=0.9\textwidth]{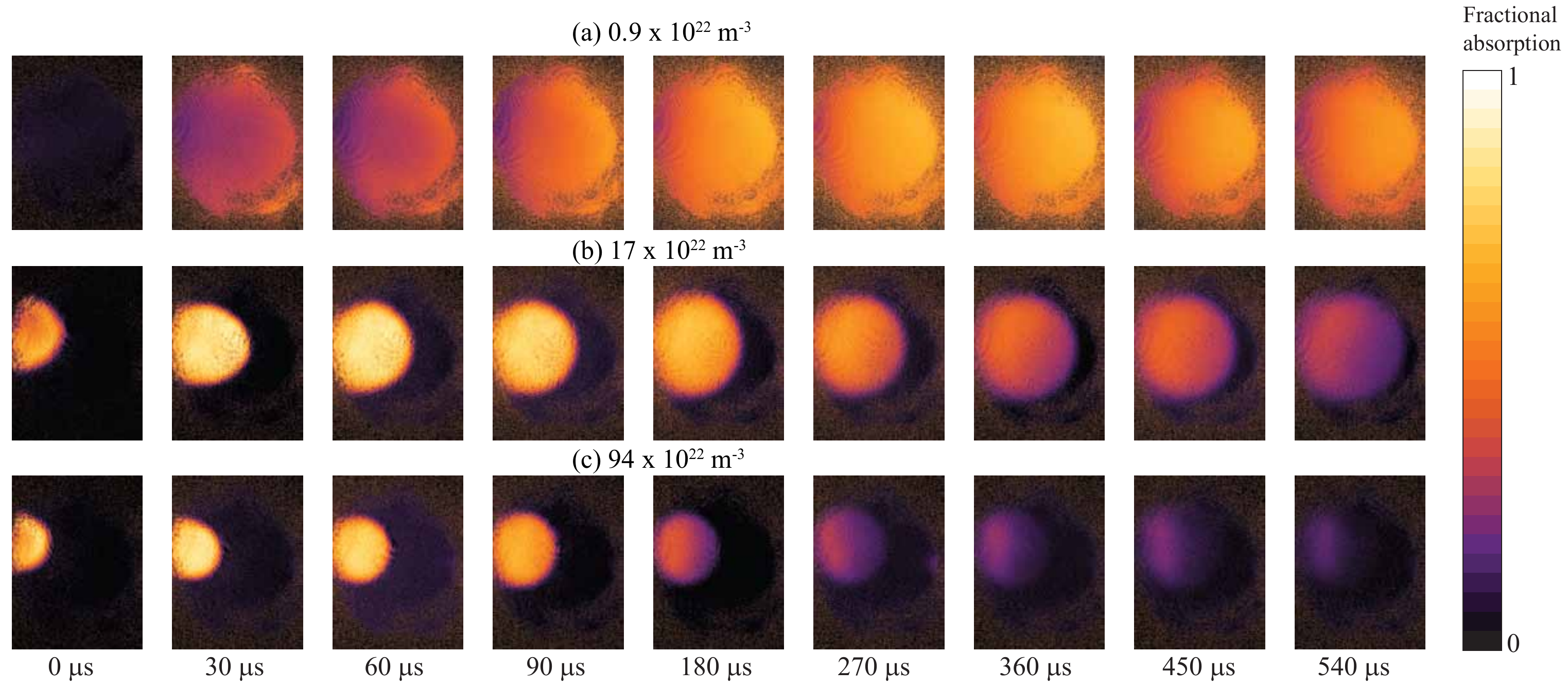}
\caption{\label{fig:images} (Color online) Absorption images of YbF at various times since ablation and for three different helium densities at a cell temperature of 20\,K.}
\end{centering}
\end{figure*}

Figure \ref{fig:images} shows a sequence of absorption images for three different buffer gas densities at a cell temperature of 20\,K. To obtain each image, we take the difference between two pictures, one where the ablation laser fires and one where it does not. The field of view is 20\,mm wide and the colour represents the fractional absorption of the probe laser. The target is on the left side of the image, just outside the field of view. The probe laser is tuned near the P-branch bandhead where there is a high density of overlapping spectral lines from rotational states in the range $N=3-6$. The camera is exposed to the probe for 10\,$\mu$s with a variable delay relative to the Q-switch of the ablation laser. The images therefore show how the density of YbF molecules in low-lying rotational states evolves with time, and how this depends on the buffer gas density.

In Fig.\,\ref{fig:images}(a), the helium density is relatively low, $n_{{\rm He}} = 0.9\times 10^{22}$\,m$^{-3}$. In this case, the molecules fill the entire field of view within 30\,$\mu$s. Their distribution is roughly uniform in all the images. The overall absorption increases for the first 180\,$\mu$s, and then slowly decreases after that. We propose that the molecules are formed close to the target, then expand ballistically, and because their mean free path is relatively long at this density, they rapidly fill the entire cell. They then diffuse slowly to the walls, where they stick. The increase in absorption over the first 180$\mu$s suggests either that the formation of new molecules continues for this length of time, or that the rotational temperature starts out high and cools on this timescale, thereby increasing the population of the low-lying rotational states being probed in the experiment. The measurements presented in Sec.\,\ref{sec:Thermalisation} confirm that the rotational temperature does indeed fall considerably over this initial period. In the first frame of the sequence of images hardly any molecules are visible because they are too hot during this initial 10\,$\mu$s period to be observed. In Fig.\,\ref{fig:images}(b) the helium density is increased to $17\times 10^{22}$\,m$^{-3}$, and the evolution has changed considerably. The initial ballistic expansion is arrested before the molecules have filled the cell, so that in the first frame we see the projection of a hemispherical distribution with a radius of about 8\,mm. This distribution then diffuses slowly into the rest of the cell. The absorption peaks after about 60\,$\mu$s, indicating that the rotational cooling is faster than before. We might expect the decay of the molecule density to be slower than before because the diffusion will be slower at higher helium density, but in fact the density appears to decay more rapidly. This is even more evident in Fig.\,\ref{fig:images}(c) where the helium density is increased to $94\times 10^{22}$\,m$^{-3}$. Here, the ballistic expansion is arrested even closer to the target, the peak absorption occurs at even earlier times, and the propagation into the rest of the cell is even slower. After 540\,$\mu$s the molecules have still not filled the entire field of view, and yet most of the molecules have vanished. The rapid decay of the signal is partly due to the proximity of the molecules to the walls of the cell. The target is close to the wall, and the molecules are stopped close to the target, so the distance that they have to diffuse before they reach a wall is small. This tends to lower their survival time in the cell, even though the mean free path is small. Other loss processes may also be contributing to the decay of the molecular density. The YbF molecules and other ablation products are confined at high density and the YbF radicals may react with these other atoms and molecules to form more stable products. We investigate the decay times of the molecules more quantitatively in the next section.

\section{Diffusion}
\label{sec:Diffusion}

We have studied the diffusion of the YbF molecules through the helium buffer gas. In order to obtain an estimate of the YbF-He diffusion cross-section, without needing an accurate measure of the helium density in the cell, we produce and probe lithium atoms in the cell simultaneously with the YbF molecules. The Li-He diffusion cross-section can be calculated accurately, and so the Li diffusion times provide a good reference against which the YbF diffusion times can be compared.

\subsection{Theory}
\label{sec:buffertheory}

Following ablation, there is an initial ballistic expansion of molecules away from the target. Over a certain distance from the target, depending inversely on the helium pressures, this ballistic expansion is arrested and the molecules then diffuse towards the walls. The flux of diffusing particles, $J$, is proportional to the gradient of the density $n$, $J=D\,\nabla n$. For diffusion of one gas into another at temperature $T$, the relationship between the diffusion coefficient, $D$, and the thermally averaged diffusion cross-section, $\bar \sigma_{\text{D}}$, is given, with sufficient accuracy, by the first Chapman-Enskog approximation \cite{Chapman:1916, Hasted:1972, Mason:1970}. Since the number density of helium, $n_{{\rm He}}$, is many orders of magnitude larger than that of our diffusing gas, the result is
\begin{equation}
D = \frac{3}{16\bar \sigma_{\text{D}} n_{{\rm He}}} \sqrt{\frac{2 \pi k_B T}{\mu}},
\label{eqn:diffusioncoeff}
\end{equation}
where $\mu=m M/(m + M)$ is the reduced mass related to the masses of the helium, $m$, and the diffusing molecules, $M$. Although this expression is only the first term of a series approximation for the diffusion coefficient, it is expected to be accurate to better than 2\% for the two cases, Li-He and YbF-He, that we consider here \cite{Mason:1970}.

The density of molecules in their journey through time and space is described by the time dependent diffusion equation $\frac{dn}{dt}=\nabla^{2}(Dn)$. The solution can be written in the form $n(r,t)=\sum_{k} c_{k} f_{k}(r)e^{-t/\tau_{k}}$,  where $c_{k}$ is the amplitude of the diffusion mode $f_{k}(r)$, which for a homogeneous diffusion coefficient is a solution of the differential equation
\begin{equation}
\frac{f_{k}}{D\tau_{k}}+\nabla^{2}f_{k}=0
\label{eqn:diffusion}
\end{equation}
subject to the boundary conditions imposed by the walls. Since the molecules stick to the walls with high probability, we take $n=0$ at the walls.  When there is no particular symmetry in the problem, three indices are needed to label the diffusion modes, represented here by the single label $k$. In all cases, the product of the diffusion coefficient and the time constant $\tau_{k}$ depends only on the cell geometry and on the indices labelling the diffusion mode. The higher order diffusion modes have smaller time constants.

If the interaction potential between the two colliding species is known, the thermally-averaged diffusion cross-section $\bar \sigma_{\text{D}}$ can be calculated. For all temperatures, $T$, of interest here, the calculation can be done classically using the following set of equations \cite{Mason:1970}:
\begin{align}
\bar \sigma_{\text{D}} &= \frac{1}{2}\int_{0}^{\infty} x^{2} e^{-x} \sigma_{\text{D}}(E)\,dx,\\
\sigma_{\text{D}}(E) &= 2\pi \int_{0}^{\infty} [1-\cos\chi(E,b)] b\,db, \\
\chi(E,b) &= \pi - 2b \int_{r_{c}}^{\infty} \frac{r^{-2}\,dr}{\sqrt{1-V(r)/E - b^{2}/r^{2}}}.
\end{align}
Here, $\sigma_{\text{D}}(E)$ is the diffusion cross-section at centre-of-mass collision energy $E$, $x=E/(k T)$, $\chi(E,b)$ is the deflection angle for a collision with energy $E$ and impact parameter $b$, $V(r)$ is the interaction potential as a function of the particle separation $r$, and $r_{c}$ is the distance of closest approach in the collision, given by $1-V(r_{c})/E - b^{2}/r_{c}^{2}=0$.

The interaction potential for Li-He is well established, there being a long history of theoretical and experimental work on this system (see \cite{Czuchaj:1995} and references therein). We have calculated $\bar \sigma_{\text{D,Li-He}}$ for a range of temperatures using the potential given in reference \cite{Czuchaj:1995}. The results at 293, 80 and 20\,K are given in Table \ref{Tab:cross-sections} on page \pageref{Tab:cross-sections}. There are two sources of error in this calculation. The first is due to our neglect of quantum effects. Using the tabulation in \cite{Mason:1970}, we estimate that quantum effects increase the cross-section by about 2\% at 20\,K, and are negligible at the higher temperatures. The second source of error is due to the uncertainty in the interaction potential. To estimate this we repeated the calculations using the alternative potential given in \cite{Patil:1991}. The cross-sections obtained from this latter potential are smaller by 0.3\% at 293\,K, 5\% at 80\,K and 6\% at 20\,K. Taking these differences as indicative of the likely accuracy, we assign a fractional uncertainty of 5\% to the cross-sections calculated at all temperatures. The interaction potential for YbF-He has also been calculated \cite{Tscherbul:2007} and we use this to obtain theoretical values for $\bar \sigma_{\text{D,YbF-He}}$ at these same temperatures. In this case, the potential is not only a function of the YbF-He separation but also of the angle between the YbF internuclear axis and the incident velocity vector of the collision. We calculate an approximate diffusion cross-section by calculating the cross-section as a function of this angle and then averaging over all angles. The results are given in Table \ref{Tab:cross-sections}.

\subsection{Model}

Since our cell geometry is not simple, we model the diffusion using finite element software\footnote{Comsol Multiphysics 3.2}. This model uses the same cell geometry as the experiment and only accounts for diffusion. It does not include the initial momentum away from the target that the molecules have following ablation. For low helium densities, this initial momentum helps to distribute the molecules throughout the cell faster than diffusion would be able to do. To understand the effect of this, we study two extreme cases for the initial condition in the model. In one case, the molecules start out with a uniform distribution over the whole cell. This is intended to model the situation at low helium density where the initial ballistic expansion completely fills the cell with molecules. In the other extreme case the initial density is taken to be $e^{-r/w}$ where $r$ is the distance away from the target and $w=1$\,mm. This is intended to model the situation at high helium density where the ballistic expansion is rapidly arrested with the molecules localized close to the target.

\begin{figure}
\includegraphics[width=0.5\textwidth]{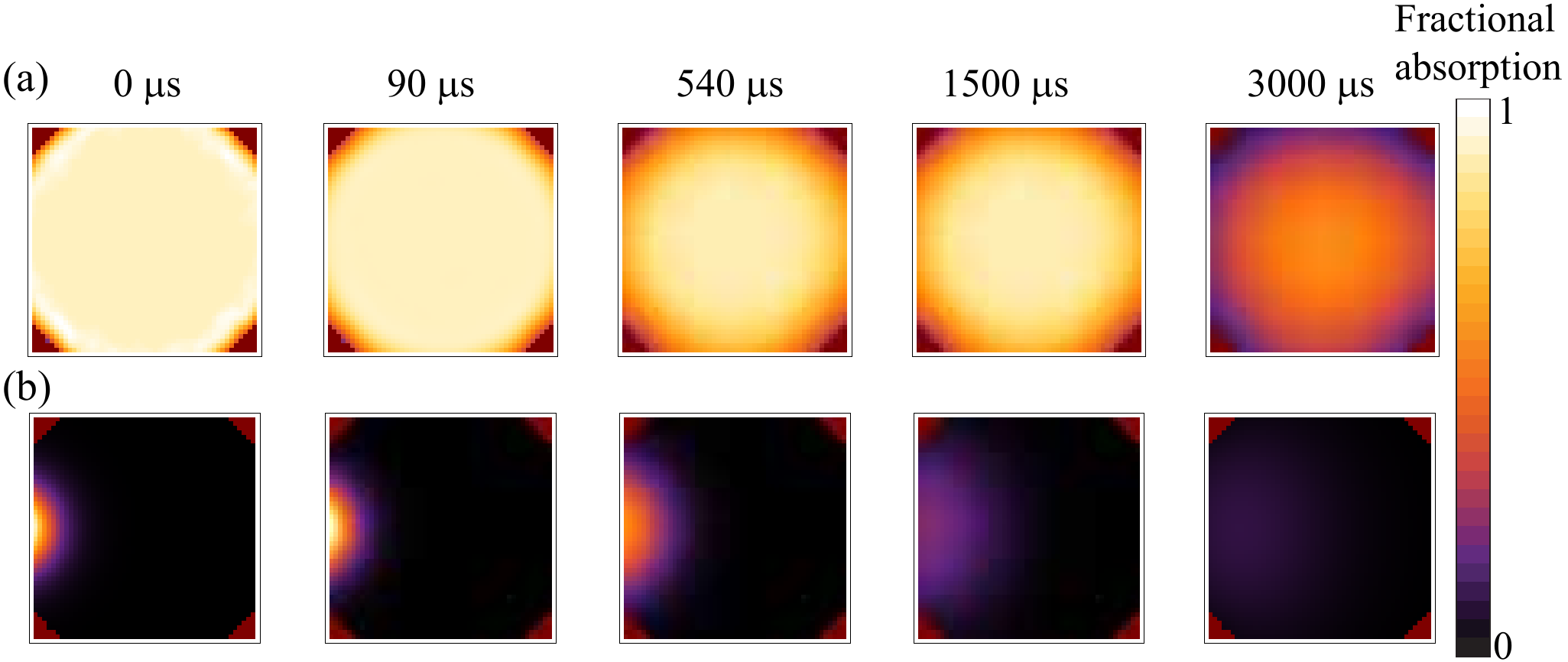}
\caption{\label{fig:simulatedImages} (Color online) Simulated absorption images for a diffusion coefficient of 0.005\,m$^{2}$\,s$^{-1}$. This corresponds, for example, to a cell temperature of 20\,K, a diffusion cross-section of $80 \times 10^{-20}$\,m$^{2}$ and a helium density of $3 \times 10^{22}$\,m$^{-3}$. (a) Uniform initial distribution throughout the cell. (b) Distribution localized near the target.}
\end{figure}

Figure \ref{fig:simulatedImages} shows the predictions of these simulations for the two extreme initial conditions. The behaviour seen in these simulated images is qualitatively similar to that of the experimental absorption images shown in Fig.\,\ref{fig:images}. In Fig.\,\ref{fig:simulatedImages}(a) the initial density distribution is a constant. This evolves to a distribution with a maximum at the middle, and that distribution then decays away slowly. In Fig.\,\ref{fig:simulatedImages}(b) the initial density is localized close to the target, and we see that as time progresses the distribution expands slowly away from the target but decays away on a faster timescale than in case (a). This is because the molecules are localized close to the target and so can diffuse back onto the target and the nearby wall. The diffusion equation tells us that high spatial gradients in density decay away rapidly. The highly localized initial distribution is represented by high-order diffusion modes localized close to the target, and these decay away quickly with a large fraction of the molecules quickly reaching the target and nearby wall, leaving only low-order modes whose amplitudes are so small that they are barely visible in the images.

\begin{figure}
\includegraphics[width=0.45\textwidth]{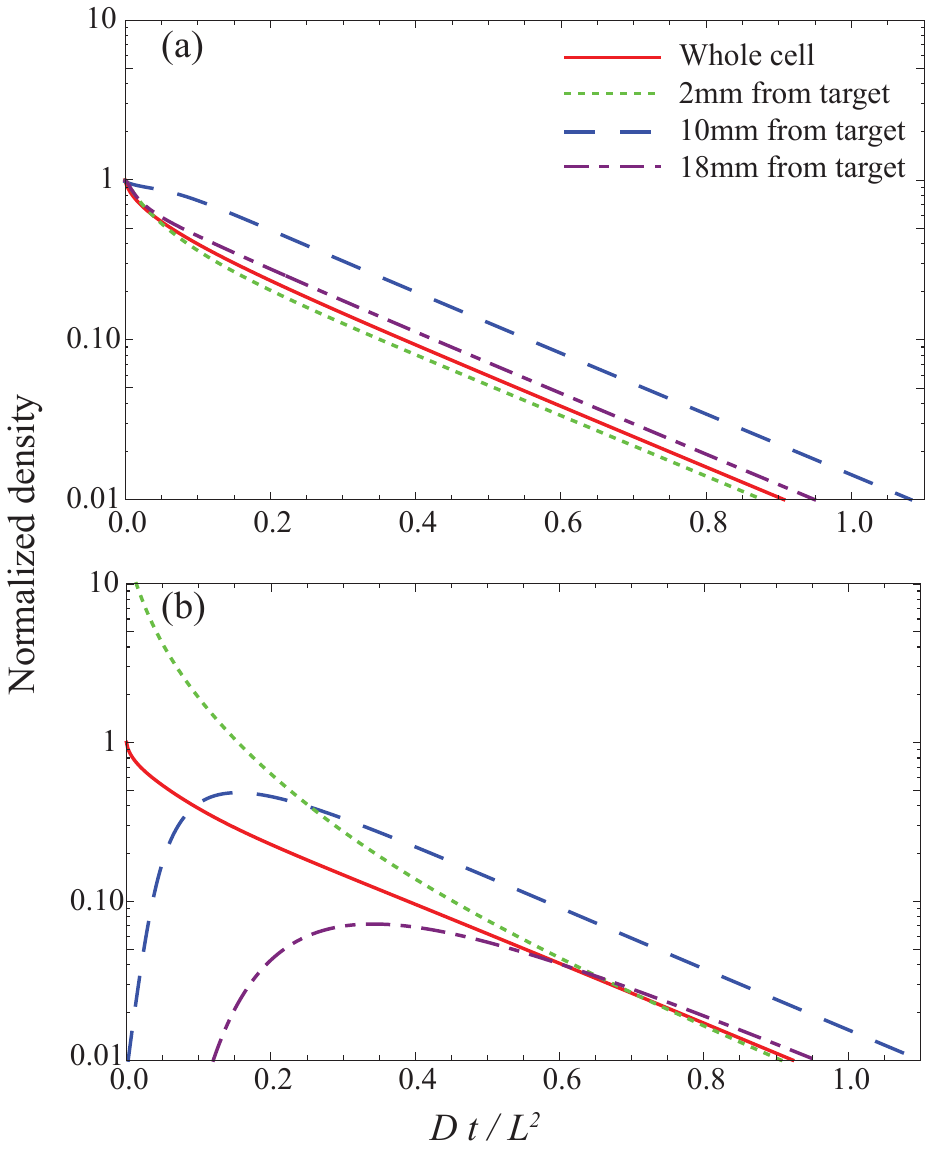}
\caption{(Color online) Simulated evolution of the molecule density at different positions in the cell. The density is normalised to the initial density averaged over the whole cell. The time axis is expressed in the dimensionless units $D t/L^{2}$, where $D$ is the diffusion coefficient and $L=1$\,cm is a characteristic size scale for the cell. (a) Uniform initial density throughout the cell. (b) Initial density localised near the target.}
\label{fig:modeldiffusiontimes}
\end{figure}

As well as taking absorption images, we measure the absorption locally using a small probe laser. To simulate these measurements, we integrate the simulated density over the volume of the probe laser. Doing this, we find that the time evolution of the density is sensitive to the position of the probe, and to the initial condition, as shown in Fig.\,\ref{fig:modeldiffusiontimes}. In this figure, we plot the simulated time evolution of the density probed at three different distances from the target, as well as the evolution of the density averaged over the whole cell. With reference to the coordinate system shown in Fig.\,\ref{fig:ExperimentSetup}, the probe laser is in the $x y$-plane, displaced along $y$. In each case, the density is normalized to the average density in the cell at $t=0$, and is plotted on a logarithmic scale. Our choice of time axis in this plot requires some explanation. Consider again the diffusion equation and introduce scaled coordinates ${\bf r}'$ related to the original coordinates $\bf{r}$ by ${\bf r}'={\bf r}/L$ where $L$ is a characteristic length scale for the cell. Its precise value is unimportant and we will take $L=1$\,cm. Taking derivatives with respect to the scaled spatial coordinates the diffusion equation is $\frac{dn}{dt'}=\nabla_{{\bf r}'}^{2}n$ where $t'= D t / L^{2}$ is a dimensionless time. This is the time axis we use in Fig.\,\ref{fig:modeldiffusiontimes} because this makes it obvious how the results scale with the overall size of the cell and with the diffusion coefficient (and hence the diffusion cross-section, helium density and temperature).

In the case where the initial density distribution is uniform [Fig.\,\ref{fig:modeldiffusiontimes}(a)], the time evolution does not depend strongly on position. There are some differences in the time evolution for different beam position at very early times, but after that the evolution is well described by a single exponential decay with a time constant which is the same at all positions. For our cell, this time constant is found to be $\tau=0.22 \times 10^{-4}/D$, where $D$ is the diffusion coefficient in m$^{2}$\,s$^{-1}$ and $\tau$ is in s. For example, for a helium density of $10^{22}$\,m$^{-3}$, a YbF-He diffusion cross-section of $80 \times 10^{-20}$\,m$^{2}$ and a temperature of 20\,K, the diffusion coefficient is $D=0.015$\,m$^{2}$\,s$^{-1}$, and the expected diffusion lifetime is $\tau=1.5$\,ms. For diffusion in a cube of side $a$, the time constant associated with the lowest order diffusion mode is $\tau=a^{2}/(3\pi^{2}D)$. It follows that the diffusion time constant for our cell is the same as that of a cubic cell with $a=2.6$\,cm. This is reassuringly similar to the size scale of our cell. The position sensitivity is far stronger when the molecules are initially concentrated near the target, as in Fig.\,\ref{fig:modeldiffusiontimes}(b). While the initial density is largest close to the source of molecules, there is a rapid decay of the density at this position. As time goes on, the rate of decay slows down, and eventually the time evolution becomes a single exponential with the same time constant as in the constant density case. However, that does not occur until the density has fallen by two orders of magnitude. Further away from the source, the density first increases with time as the molecules arrive at the probe, and then it decreases again with the same time constant as in the constant density case. The peak density decreases as we move further from the target, and the time taken to reach this peak increases with increasing distance. We conclude that at long times the time evolution is insensitive to either the initial condition or the position of the probe, but that at short times it is sensitive to probe position particularly when the initial density is strongly localised.

These observations can be explained by expressing the density distribution as a sum over a set of diffusion modes, the higher order modes representing more rapid spatial variations in density and decaying away more quickly. When the cell is uniformly filled, the density rapidly settles into the lowest order diffusion mode, which gives us the longest diffusion time. To represent an initial distribution that is strongly localized we need to put large amplitude into high-order modes, and these then decay away rapidly. Eventually, only the low-order modes, which persist for much longer, will remain, but the amplitude of the low-order mode may be very small. In an experiment, it may be too small to observe above the noise.

\subsection{Experimental results}

We compare the YbF diffusion times to the diffusion time of lithium atoms introduced into the cell at the same time. If we assume that the initial distribution is the same for the YbF and the Li, the product of the diffusion coefficient and the time constant is the same for both species, since this product is only a function of the cell geometry. It follows that
\begin{equation}
\bar \sigma_{D,2}=\bar \sigma_{D,1} \sqrt{\frac{\mu_1}{\mu_2}} \frac{\tau_2}{\tau_1}
\label{Eq:sigma1sigma2}
\end{equation}
where subscript 1 refers to Li-He and subscript 2 to YbF-He. To introduce Li into the cell, a piece of lithium wire was mounted into the target holder along with the AlF$_3$/Yb target. The two materials were ablated simultaneously using a single Nd:YAG laser spot. The lithium density was monitored by absorption of light from a diode laser tuned into resonance with the $^{7}$Li D1 line at 671\,nm. This probe laser was overlapped with the YbF probe beam and the absorption signals for the two species were measured simultaneously on two separate photodiodes. In this way, both species were measured at the same place and at the same time, and both were produced in the same place on the target. The probe lasers were in the $x y$-plane (see Fig.\,\ref{fig:ExperimentSetup}), about 7\,mm from the target. It was possible to obtain extremely high optical depths for Li, but through judicious positioning of the ablation spot on the target we were able to obtain optical depths of the same order of magnitude for the two species. Because the Li and YbF densities are many orders of magnitude lower than the He density collisions between them are not important.

\begin{figure}
\includegraphics[width=0.45\textwidth]{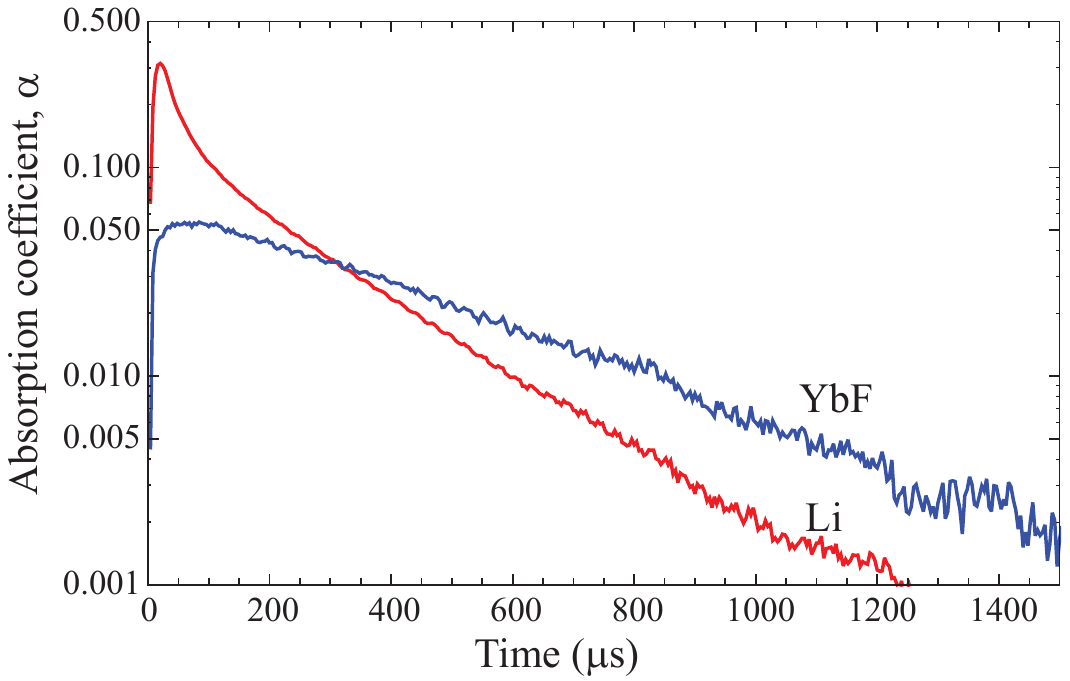}
\caption{(Color online) Absorption coefficient on a logarithmic scale plotted as a function of time for Li (red) and YbF (blue) at a helium density of $2.2 \times 10^{22}$\,m$^{-3}$ and a cell temperature of 293\,K.}
\label{fig:timeEvolution}
\end{figure}

Figure \ref{fig:timeEvolution} shows the time evolution of the YbF and Li absorption signals for a helium density of $2.2 \times 10^{22}$\,m$^{-3}$. At this density we expect the molecules to fill the entire cell through ballistic expansion. We plot the absorption coefficient, $\alpha$, which is directly proportional to the molecule density, the interaction length and the absorption cross-section, and is related to the fractional absorption ${\cal A}$ via $\alpha = - \ln \left(1-{\cal A}\right)$ (see appendix \ref{App1}). Transients due to optical pumping and collisional redistribution of the population are too fast to be observable in the experiment. The YbF density is probed near the A-X(0-0) bandhead where there is a dense cluster of overlapping P-branch lines. For Li, there is an initial fast rise in the density during the first 20\,$\mu$s as the atoms fill the cell. After that, the signal is described well by a double exponential decay, one exponential having a short time constant that dominates during the first 150\,$\mu$s, and the second having a longer time constant that dominates at later times. This same behaviour is observed over a wide range of buffer gas densities. We find that as the helium density increases the short time constant decreases from about 40\,$\mu$s when the density is $0.25 \times 10^{22}$\,m$^{-3}$ to about 20\,$\mu$s when the density is $2.5 \times 10^{22}$\,m$^{-3}$. We suppose that this fast decay is the decay of higher-order diffusion modes, the result being sensitive to the initial condition and the position of the probe laser as discussed above. The long time constant on the other hand increases with buffer gas density, and we take this to be the characteristic diffusion time for Li in the cell. The YbF absorption curves are influenced by both diffusion and by rotational cooling which, at early times, increases the relative population in the low-lying rotational states probed in the experiment, particularly for low cell temperatures. In Sec.\,\ref{sec:Thermalisation} we measure how the rotational temperature changes with time. A model that includes both diffusion and the rotational cooling that we have measured predicts that, for a cell temperature of 20\,K, the change in absorption is strongly influenced by rotational cooling at early times, but is dominated by diffusion at later times. At higher temperatures the influence of rotational cooling is far weaker. To obtain the YbF diffusion time constants we fit single exponential decays to the late part of the absorption curves ($t > 0.32$\,ms for the 20\,K data and $t>0.14$\,ms for 293\,K). Our model shows that we will obtain accurate diffusion times using this procedure. We have also verified that, at 20\,K, fitting only to data beyond 0.6\,ms makes a negligible difference to our results.

\begin{figure}
\includegraphics[width=0.45\textwidth]{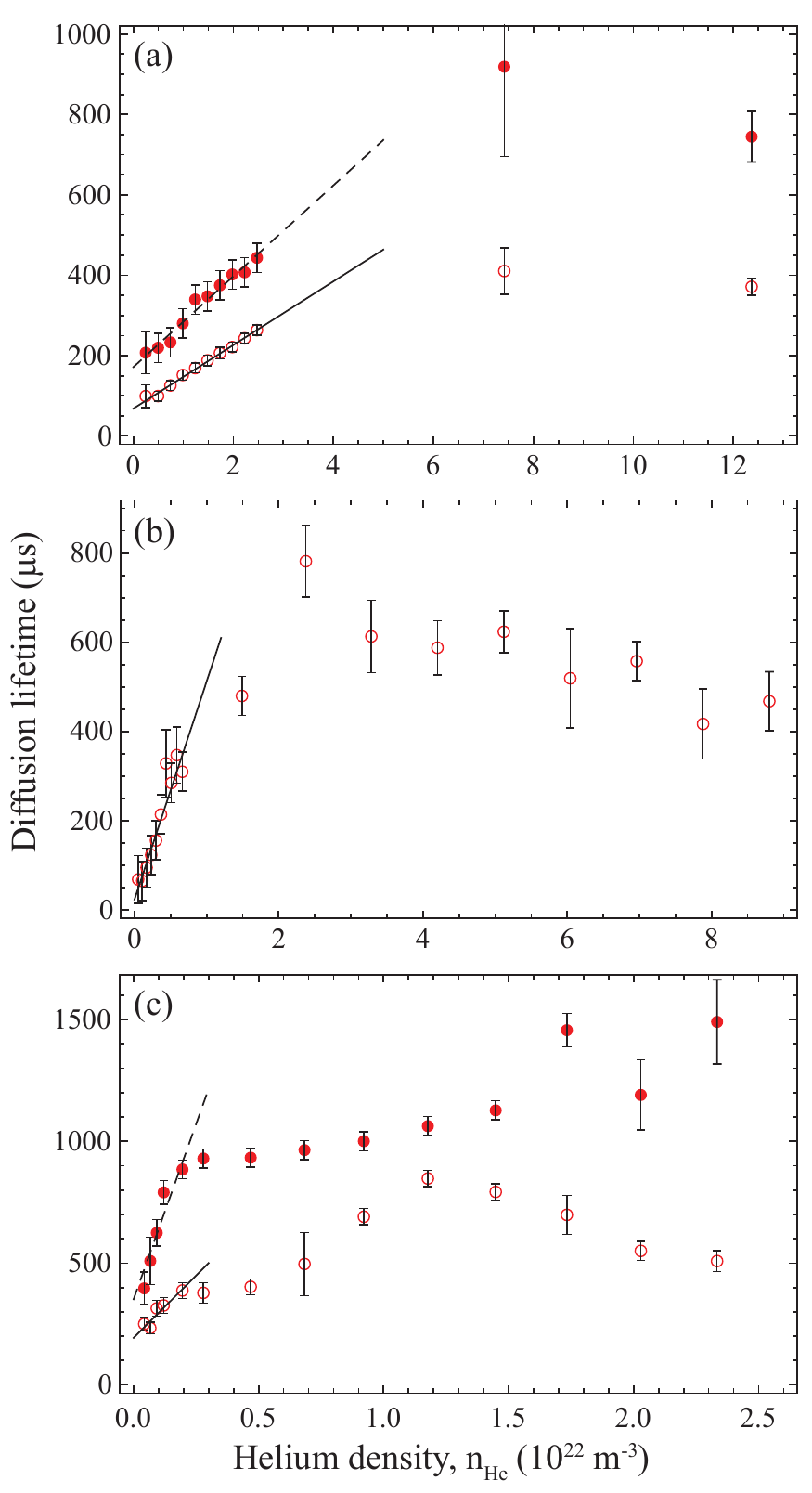}
\caption{(Color online) Diffusion time constants as a function of buffer gas density for Li (open circles) and YbF (filled circles) for buffer gas temperatures of (a) 293\,K, (b) 80\,K (Li only) and (c) 20\,K. The lines (solid for Li, dashed for YbF) are linear fits to the data at low buffer gas density. Note the different scales on both axes for the three graphs.}
\label{fig:experimentalDiffusionTimes}
\end{figure}

Figure \ref{fig:experimentalDiffusionTimes} shows the characteristic diffusion time constants for Li and YbF as a function of helium density, for cell temperatures of 293\,K, 80\,K and 20\,K. At 80\,K no useful YbF data was obtained, so only Li data are shown. The error bars are determined from the scatter of repeated measurements taken under the same nominal conditions. They therefore appear as errors in the time constant though we believe the major source of the scatter to be due to the difficulty of setting the buffer gas density, which was very slow to reach equilibrium particularly at low temperatures. For low buffer gas densities, all the measured time constants increase linearly with density and both species diffuse more slowly at lower temperatures as expected in a simple model of diffusion. At both 293\,K and 20\,K the time constant for YbF increases more rapidly with increasing helium density than for Li. At higher densities however, the diffusion times stop increasing and become roughly independent of the helium density. This behaviour is displayed by both YbF and Li at 293\,K, by Li at 80\,K and YbF at 20\,K. The data for Li at 20\,K appear to show a more complicated behaviour at high density. For Li, the lifetimes level off at approximately 400\,$\mu$s at 293\,K, 600\,$\mu$s at 80\,K, and roughly 500\,$\mu$s at 20\,K. For YbF, the corresponding values are approximately 800\,$\mu$s at 293\,K and 1000\,$\mu$s at 20\,K. The density at which the levelling off occurs is about the same for the two species, and shifts to lower values as the temperature is lowered.

At low density, the diffusion times show the expected linear increase with helium density, so we fit straight lines to the low density parts of the five datasets. The fit results are shown by the lines in Fig.\,\ref{fig:experimentalDiffusionTimes}. For the 293\,K and 80\,K data, the first 10 data points were used for the linear fit. For the 20\,K data it is unclear whether to fit to the first 4 or first 5 data points (the latter is shown in Fig.\,\ref{fig:experimentalDiffusionTimes}(c)), and so we do both. The gradients of these two fits are consistent within errors, and we take the mean value as our best estimate. The gradients of the linear fits, together with Eq.\,(\ref{Eq:sigma1sigma2}) and the calculated Li-He diffusion cross-section, provide a measurement of the YbF-He diffusion cross-section, $\bar \sigma_{\text{D,YbF-He}}$. The results are given in the fourth column of Table \ref{Tab:cross-sections}. The uncertainty is derived by adding in quadrature the statistical errors in the gradients of the linear fits, the error in the Li-He cross-section estimated above, and in the case of the 20\,K result, half the difference between the gradients of the 4-point and 5-point fits. Neither the helium density or the cell temperature enter in the determination of the cross-section. Errors arising due to a difference in the initial spatial distribution of the Li and YbF or small displacements of the two probes are deemed to be negligible because at these helium densities the initial ballistic expansion fills the cell, as in Fig.\,\ref{fig:images}(a), and our simulation results indicate that the diffusion time is insensitive to the exact initial distribution or probe position in this case. Errors arising due to a difference in temperature between the two species are also negligible because the time for thermalization with the helium is much smaller than the characteristic diffusion time (see Sec.\,\ref{sec:Thermalisation}). Our experimental values for the YbF-He diffusion cross-section are consistent with our theoretical values. In particular, the values are in agreement at 293\,K where the experimental uncertainty is smallest.

\begin{table*}
\begin{ruledtabular}
\begin{tabular}{cccccc}
T & $\bar \sigma_{\text{D,Li-He}}$ (theory) & $\bar \sigma_{\text{D,YbF-He}}$ (theory) & $\bar \sigma_{\text{D,YbF-He}}$ (experiment) & $\tau_{\text{Li}}/n_{\text{He}} $ (theory) & $\tau_{\text{Li}}/n_{\text{He}}$ (experiment) \\
(K) & ($10^{-20}$m$^{2}$) & ($10^{-20}$m$^{2}$) & ($10^{-20}$m$^{2}$) & ($10^{-22}\mu$s/m$^{-3}$) & ($10^{-22}\mu$s/m$^{-3}$) \\
\hline
293 & 35.8 & 40.9 &$41 \pm 4$ & 171 & $79 \pm 2$ \\
80 & 51.9 & 55.8 & - & 482 & $489 \pm 48$ \\
20 & 71.0 & 79.6 & $203 \pm 75$ & 1300 & $1022 \pm 226$ \\
\end{tabular}
\end{ruledtabular}
\caption{
\label{Tab:cross-sections} Diffusion cross-sections and lifetimes for Li and YbF at three temperatures. The second and third columns give theoretical values for the Li-He and YbF-He diffusion cross-sections, calculated using the potentials of \cite{Czuchaj:1995} and \cite{Tscherbul:2007} respectively. The fourth column gives experimental values for the YbF-He diffusion cross-section obtained from the data shown in Fig.\,\ref{fig:experimentalDiffusionTimes}, calibrated by the Li-He cross-sections. The fifth and sixth columns give theoretical and experimental values for the Li diffusion lifetime divided by the helium density. The theoretical values are determined from the theoretical cross-sections and a numerical model of diffusion in the cell. The experimental values are the slopes of the straight line fits shown in Fig.\,\ref{fig:experimentalDiffusionTimes}.}
\end{table*}

Extrapolating the data to zero helium density as shown by the lines in Fig.\,\ref{fig:experimentalDiffusionTimes} we observe an offset in all of the lifetimes. This suggests that the ablation process itself puts a significant density of material in the cell which the YbF and Li then have to diffuse through. This is consistent with the offset observed in Fig.\,\ref{fig:pgraph}. The offset is smaller for the Li data at 80\,K. We do not know the reason for this. We speculate that, when the cell is warm, vacuum contaminants such as water and hydrocarbons reach the cell and are absorbed into the target, and when the cell is very cold the target absorbs helium just like a cryo-sorb. At 80\,K neither occur because this is not cold enough to absorb helium but is too cold for water and hydrocarbons to reach the cell.

\subsection{Discussion}

We have used the Li data to calibrate the YbF data and so our determination of the YbF-He diffusion cross-section does not rely on the diffusion model or on knowing the helium density. It is nevertheless interesting to compare the experimental results at low density with the predictions of the diffusion model for Li. We assume that at low helium densities, the cell is immediately filled with a uniform density of particles. Fitting a sum of two exponentials to the results of the model with this initial condition gives the longer of the two time constants as $\tau=0.22 \times 10^{-4}/D$, where $D$ is the diffusion coefficient in m$^{2}$\,s$^{-1}$ and $\tau$ is in s. As noted above, the long time constant is insensitive to the initial condition, so the error arising from the uncertainty in the initial distribution is small. Using the definition of $D$ given in Eq.\,(\ref{eqn:diffusioncoeff}) and the calculation of the Li-He diffusion cross-sections, we obtain a theoretical prediction for the gradient $\tau_{\text{Li}}/n_{\text{He}}$. The theoretical results obtained at 293, 80 and 20\,K are given in Table \ref{Tab:cross-sections} where they are compared with the experimental gradients obtained from the fits shown in Fig.\,\ref{fig:experimentalDiffusionTimes}. The error bars given here are the fitting errors, and do not include a possible systematic error in the determination of the helium density. The experimental and theoretical results are in good agreement at 80\,K and 20\,K, but disagree by a factor of approximately 2 at 293\,K. We do not have an explanation for this discrepancy. The temperature only enters the model through the diffusion coefficient, so if the discrepancy is due to an error in the model it must be that the calculated diffusion cross-section is accurate at 80\,K and 20\,K, but not at 293\,K. This seems unlikely. The alternative is that there is a systematic error in the experiment which is large at 293\,K, but small at the lower temperatures. The most obvious source of systematic error is in the determination of the helium density, but here we would expect the error to be smallest at 293\,K since the measurements were carefully calibrated at this temperature. We note that it is quite common to determine diffusion cross-sections from diffusion times in a buffer gas cell. To do so without using a reference atom requires a careful calibration of the buffer gas density and, unless the shape of the cell is particularly simple, a numerical model of diffusion in the actual cell geometry being used.

Next, we turn to the question of why the diffusion times flatten off at higher helium density. A similar behaviour, suggestive of an additional loss mechanism, has been noted in several other papers e.g. \cite{Weinstein(1):1998, Sushkov:2008, Lu:2008, Lu:2009}. In reference \cite{Sushkov:2008} the decay times were measured as a function of helium density for lithium, rubidium, silver and gold. In all cases, the decay times first increased linearly with helium density, but then turned over and gradually decreased with increasing density. The authors suggest three possible loss channels that might be responsible for the shortening of the lifetime at high density: the formation of dimers, capture of atoms by clusters formed during the ablation process, or atom loss on impurities present in the buffer gas or ablated off the target. One of these mechanisms might be responsible for the behaviour we observe. Here, we suggest a fourth mechanism, based only on diffusion and on the way the initial molecule distribution depends on helium density. The absorption images, Fig.\,\ref{fig:images}, show that there is a rapid initial ballistic expansion of molecules into the cell which sets the initial condition for the subsequent diffusion to the walls. When the density is low, the molecules rapidly fill the entire cell and so the initial distribution is set by the cell geometry and is independent of density. In this regime the diffusion time constant is proportional to the helium density. At higher densities however, an initial cloud of molecules is produced whose size depends on the helium density rather than on the cell size. The higher the density, the smaller the size of the initial cloud. This localization of the cloud results in a more rapid initial decay, as indicated in Fig.\,\ref{fig:modeldiffusiontimes}, particularly if the probe is moved to a place where the signal is initially high, which it is natural to do experimentally. We see that at high density there are two competing density-dependent effects: increasing the density decreases the diffusion coefficient which slows down the diffusion, but increasing the density also reduces the initial size of the distribution which in turn drives faster diffusion, at least at the beginning. The turn-over from the proportional to the weakly-dependent regime occurs once the initial ballistic expansion fails to fill the cell, and this will depend on the cell's size and on the energy of the ablation pulse. Ideally, it would be possible to observe the multi-exponential decay that will occur at high density, and to find the longest decay time which will be independent of the initial condition. In practice the amplitude of the lowest order mode may be too small to observe, in which case the initial condition will inevitably affect the measurement of the time constant in the way we have described.

\section{Thermalization}
\label{sec:Thermalisation}

The translational and rotational temperatures can be determined by recording Doppler-broadened spectra of the YbF molecules. The translational temperature is found from the Gaussian widths of the lines, and the rotational temperature is determined from the relative strengths of transitions originating from different rotational states. Since we record the time-evolution of the absorption at each laser frequency, we can determine the translational and rotational temperatures as a function of time.

\subsection{Translational temperature}

\begin{figure}
\begin{center}
\includegraphics[width=0.45\textwidth]{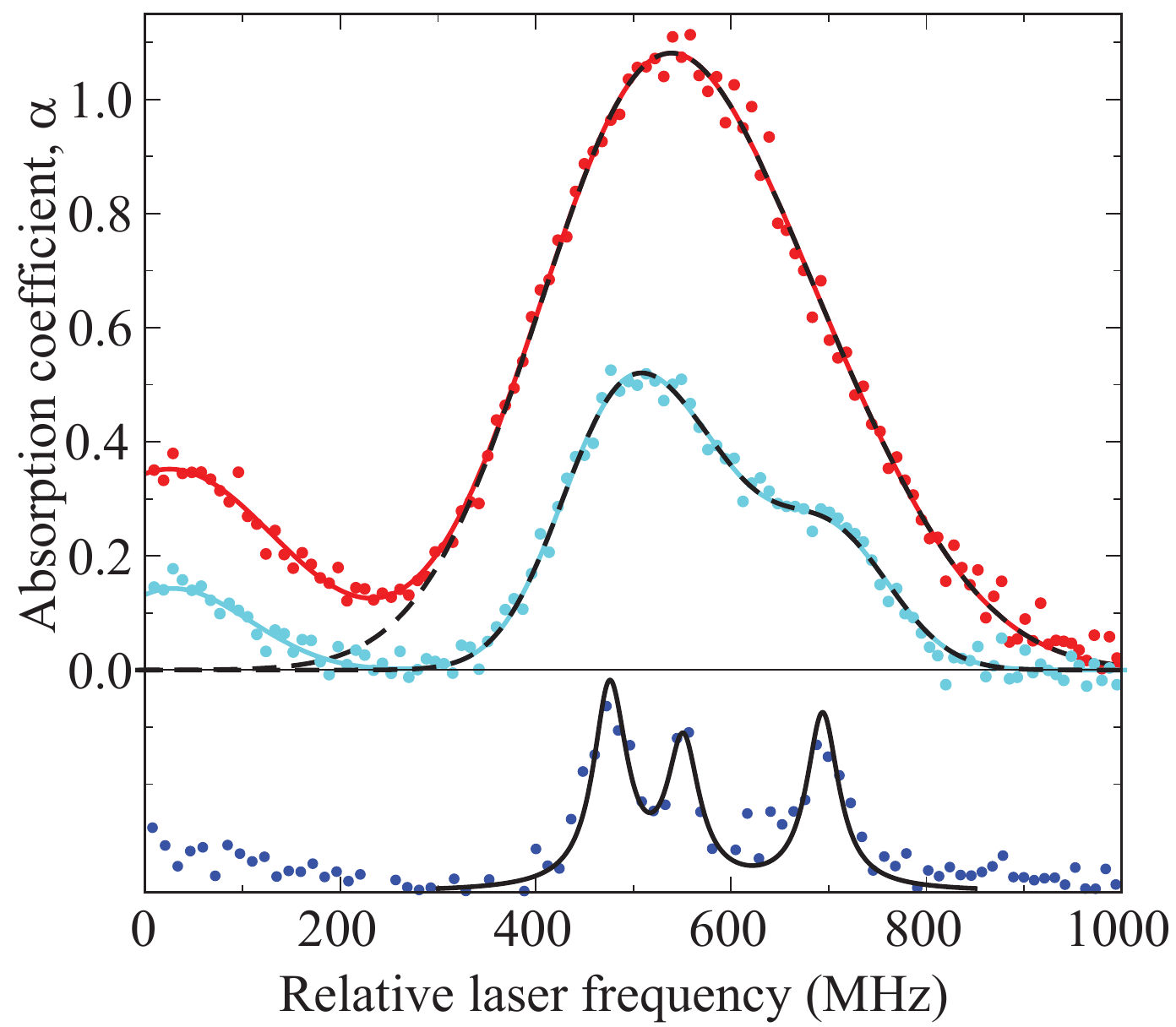}
\caption{\label{fig:dopplerfits}
(Color online) Upper traces: Doppler-broadened absorption spectra of the Q(7) transition with a cell temperature of approximately 20\,K. Various hyperfine components and YbF isotopologues contribute to the absorption. The dark red markers correspond to a spectrum recorded 200\,$\mu$s after the ablation pulse that produces the molecules. The light blue markers show data recorded 2000\,$\mu$s after ablation. The solid lines are multi-component Voigt fits to the data. The dashed lines isolate the components in the fit corresponding to the $^{174}$YbF isotopologue. Lower trace: Saturated absorption spectrum recorded simultaneously allowing the line centres of the hyperfine components to be determined and fixed in the fits to the Doppler-broadened spectra.
}
\end{center}
\end{figure}

The measurement of the width and the area of the Doppler-broadened lines is complicated by hyperfine splittings that are comparable to the Doppler widths of the features themselves, and by the existence of several YbF isotopic species. The use of Doppler-free saturated absorption spectroscopy enables us to identify the centre frequencies of the underlying components, and hence to fit a combination of multiple Voigt profiles to the Doppler-broadened spectra \cite{Skoff:2009}. Figure~\ref{fig:dopplerfits} shows typical Doppler-broadened spectra in an experiment where the cell temperature was approximately 20\,K and the helium density was approximately $0.9\times10^{22}$\,m$^{-3}$. Two spectra are shown, one averaged over a 200\,$\mu$s wide time-window centred 200\,$\mu$s after the ablation pulse, and the other averaged over a window of the same width but centred 2000\,$\mu$s after the the ablation pulse. In the early time window the presence of multiple hyperfine components is hidden by the large Doppler width, but this is revealed in the later time window when the translational temperature is lower. Underneath the Doppler-broadened profiles is shown the saturated absorption spectrum that was recorded simultaneously. There are four hyperfine components, though two of them remain unresolved in the Doppler-free spectrum so we only see three resolved spectral lines. Knowing the spacing of the hyperfine components from the Doppler-free spectrum, we fit to the Doppler-broadened spectra using a sum of Voigt profiles, where the only free parameters are a single Gaussian width and the amplitudes of the various components. The width of the Lorentzian in the Voigt profile is fixed at 30\,MHz which is the typical width obtained from fitting Lorentzians to a large number of lines in the Doppler-free spectrum. The fitted widths give translational temperatures of 76$\pm$4K and 23$\pm$1K for the two sets of data shown, clearly indicating the translational cooling of the YbF towards the temperature of the cell walls.

\subsection{Rotational temperature}

\begin{figure}
\begin{center}
\includegraphics[width=0.4\textwidth]{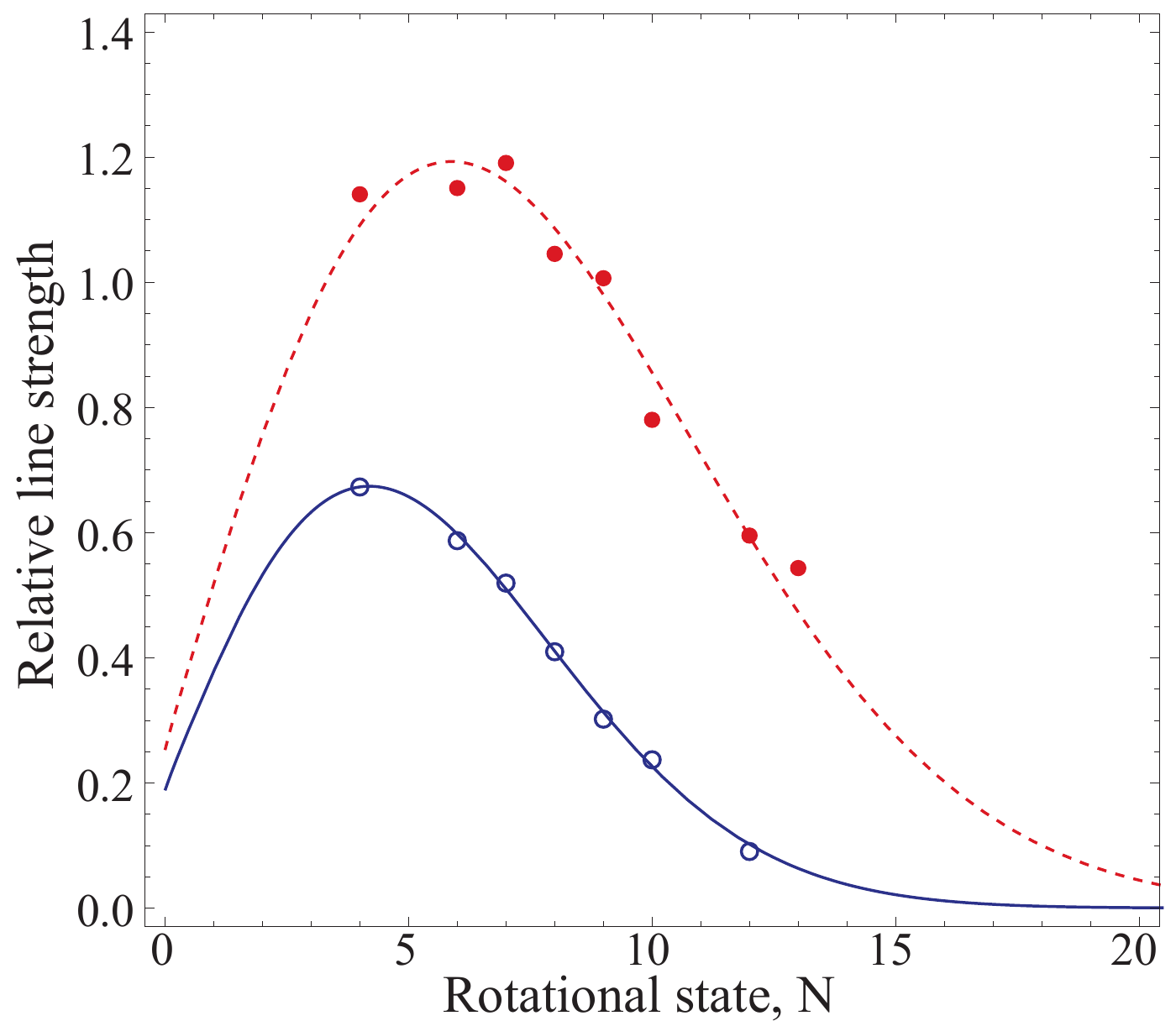}
\caption{\label{fig:rotationalstates}
(Color online) Relative strengths of Q-branch transitions originating from different rotational states $N$, at 400\,$\mu$s after ablation (filled points) and 2000\,$\mu$s after ablation (open points). The line strengths are determined from the areas of the Doppler-broadened spectral lines. The cell temperature was approximately 20\,K and the helium density approximately $0.9\times10^{22}$\,m$^{-3}$. The solid and dashed lines are fits to the data, with the rotational temperature as a free parameter in the fit.}
\end{center}
\end{figure}

The rotational temperature of the molecules is determined from the relative strengths of absorption lines originating from different rotational states $N$. We obtain the relative line strengths from the areas of the Voigt profiles fitted to the Doppler-broadened spectra. Figure \ref{fig:rotationalstates} shows the relative line strengths of the Q-branch transitions with $N$ between 4 and 13, determined at two different times, 400\,$\mu$s and 2000\,$\mu$s after the ablation pulse. The width of the time window is 200\,$\mu$s in both cases. The fits shown in the figure are Boltzmann distributions, modified by the slight dependence of the transition matrix elements on the rotational state. These fits determine the rotational temperature to be 37$\pm$3\,K at 400\,$\mu$s and 20$\pm$1\,K at 2000\,$\mu$s indicating that the rotational temperature of the YbF also cools over this timescale.

\subsection{Thermalization models}

\begin{figure}
\begin{center}
\includegraphics[width=0.45\textwidth]{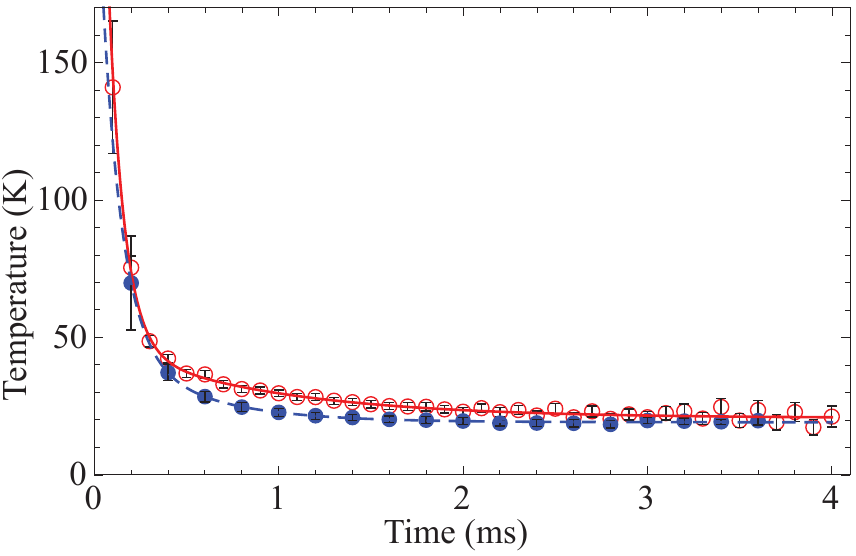}
\caption{\label{fig:temperatureplot}
(Color online) Measured temperatures of the translational motion (open points, red) and rotational motion (filled points, blue) when the buffer gas cell is maintained at approximately 20\,K and the helium density is approximately $0.9\times10^{22}$\,m$^{-3}$. Some of the error bars are smaller than the size of the data points. The solid and dashed lines are double exponential fits to the translational and rotational data, as described in the text.
}
\end{center}
\end{figure}

Figure \ref{fig:temperatureplot} shows the evolution of the translational and rotational temperatures plotted at intervals of 100 and 200\,$\mu$s, respectively. The error bars indicate the uncertainties in the parameters obtained from fits such as those shown in figures~\ref{fig:dopplerfits} and~\ref{fig:rotationalstates}. The figure shows that the translational and rotational degrees of freedom are in close equilibrium throughout the cooling process. The data also indicate that the molecules initially cool rapidly during the first 200\,$\mu$s, and then continue to cool but on a much longer time scale.

We fit this data to a model in which the temperature, $T$, of the molecules evolves with time according to
\begin{equation}
T = T_{c} + T_{i1}e^{-t/\tau_{1}} + T_{i2}e^{-t/\tau_{2}}.
\label{Eq:DoubleExponentialThermalisationModel}
\end{equation}
This model fits very well to our data, as indicated by the best fit lines shown in Fig.\,\ref{fig:temperatureplot}. Indeed, the quality of the fits to the data tend to be better than would be expected if the error bars were purely statistical. This is because the data at different times are obtained from the same experimental sequence, with the absorption integrated over different time windows, and so there is a high degree of correlation in the noise at different times. For the translational motion the best fit parameters are $T_{c}=21\pm1$\,K, $T_{i1} = 329\pm117$\,K, $T_{i2}=28\pm3$\,K, $\tau_{1}=84\pm14$\,$\mu$s and $\tau_{2}=880\pm160$\,$\mu$s. For the rotational motion, the fit parameters are less precise but are all consistent with the above values.

We now explain why we choose to fit to this model, and we interpret the two time constants. In a simple thermalization model \cite{deCarvalho:1999}, the temperature of the molecules is given by
\begin{equation}
\frac{d T}{d t} = -R(T- T_{\mathrm{bg}})/\kappa,
\end{equation}
where $R$ is the collision rate, $T_{\mathrm{bg}}$ is the buffer gas temperature and $\kappa=(M+m)^{2}/(2 M m)$. The collision rate is given by the product of the helium density, $n_{\text{He}}$, the collision cross-section, $\sigma$, and the thermally-averaged relative speed $\bar{v}$, $R=n_{\text{He}}\sigma\bar{v}$. The helium atoms are much lighter than the YbF molecules and so, over the temperature range of interest here, $\bar{v}$ is dominated by the speed of the helium and has little dependence on the temperature of the molecules. In this case, provided the collision cross-section depends only weakly on the temperature, the collision rate is approximately constant. Then the solution is simply
\begin{equation}
T=(T_0 - T_{\mathrm{bg}}) \exp[-R t/\kappa] +T_{\mathrm{bg}},
\label{Eq:temperatureEvolution1}
\end{equation}
where $T_{0}$ is the initial temperature of the molecules. This single exponential decay does not fit well to the data. Modifying this model to take into account the dependence of $\bar{v}$ on the molecule temperature does not improve the fit.

This leads us to propose a more complicated model in which the molecules thermalise with the buffer gas as before, but the buffer gas temperature is initially raised by the ablation process to $T_{\mathrm{bg,0}}$, and then cools with a time constant $\tau_{\mathrm{bg}}$ to the temperature of the cell walls, $T_{\mathrm{c}}$. The function that describes this model is the same as Eq.\,(\ref{Eq:temperatureEvolution1}), but with the buffer gas temperature now a function of time, $T_{\mathrm{bg}}=(T_{\mathrm{bg,0}}-T_{\mathrm{c}}).\exp[-t/\tau_{\mathrm{bg}}] + T_{\mathrm{c}}$. This model fits the data very well. For the translational motion, this fit suggests that the cross-section for thermalization of the YbF with the helium is $\sigma = 7.1 \pm 1.0 \times 10^{-20}$m$^{2}$, that the buffer gas is initially at $T_{\mathrm{bg,0}}=48\pm$4\,K and that it cools to $T_{\mathrm{c}}=20\pm$1\,K with a time constant of $\tau_{\mathrm{bg}}=915\pm$180\,$\mu$s. Fitting the same model to the rotational temperature data, we obtain similar fit parameters:  $\sigma = 6.8 \pm 6.0\times 10^{-20}$m$^{2}$, $T_{\mathrm{bg,0}}=49\pm$36\,K, $T_{\mathrm{c}}=19\pm$1\,K, and $\tau_{\mathrm{bg}}=460\pm$250\,$\mu$s. The energy required to explain the 30K increase in buffer gas temperature corresponds to roughly 0.5\% of the pulse energy of the ablation laser being rapidly transferred to the helium buffer gas. A time constant of about 1\,ms for re-thermalization of the buffer gas with the cold walls of the cell is as expected given the size of the cell and the thermal diffusivity of helium at these temperatures and pressures. However, an atom-molecule thermalization cross-section of about $7 \times 10^{-20}$m$^{2}$ is smaller than we would expect in light of the measurements presented in Sec.\,\ref{sec:Diffusion}. It also seems unlikely that the time constant for rotational thermalization would be so similar to that of translational thermalization. Therefore, although the model fits to the data, it only does so with parameters that we find unrealistic.

These observations suggest that the rotational and translational motions thermalize with the helium on a timescale that is shorter than the time resolution of the experiment, and the temperature evolution we observe in Fig.\,\ref{fig:temperatureplot} is entirely that of the helium. This implies that the heat of ablation deposited into the helium is dissipated with two different time constants, one being about 10 times longer than the other. We suppose that the heat is initially deposited into a localized region close to the target. This hot helium then mixes with the cold helium so that, on the timescale $\tau_{1}$, the temperature becomes roughly uniform throughout the cell, but higher than that of the cell walls. Finally, the entire bulk of helium cools to the cell temperature with a longer time constant $\tau_{2}$. In terms of diffusion modes, the initial temperature distribution is represented by a sum of modes, the higher-order ones representing the more rapid variations in temperature and decaying away rapidly, leaving a small amplitude lowest-order mode which decays slowly. This model for the diffusion of heat is exactly analogous to the diffusion of the molecules in the cell, giving the double exponential decay that we have fit to the data, Eq.\,(\ref{Eq:DoubleExponentialThermalisationModel}). Modelling the diffusion of heat to the cell walls using the thermal diffusivity of helium at this temperature and density, we find that the time constants found from the fit are reasonable. We conclude that this is the most likely explanation for the cooling we observe.

\section{Molecule density and saturation of the absorption}

All of the measurements presented in this paper are based on absorption of a probe laser. In this section, and in Appendix \ref{App1}, we consider this absorption in more detail. We are particularly concerned to obtain an accurate value for the molecule density from the observed absorption, and to understand how the fractional absorption depends on the probe intensity and on the collision rates between the molecules and the helium.

\begin{figure}
\includegraphics[width=0.45\textwidth]{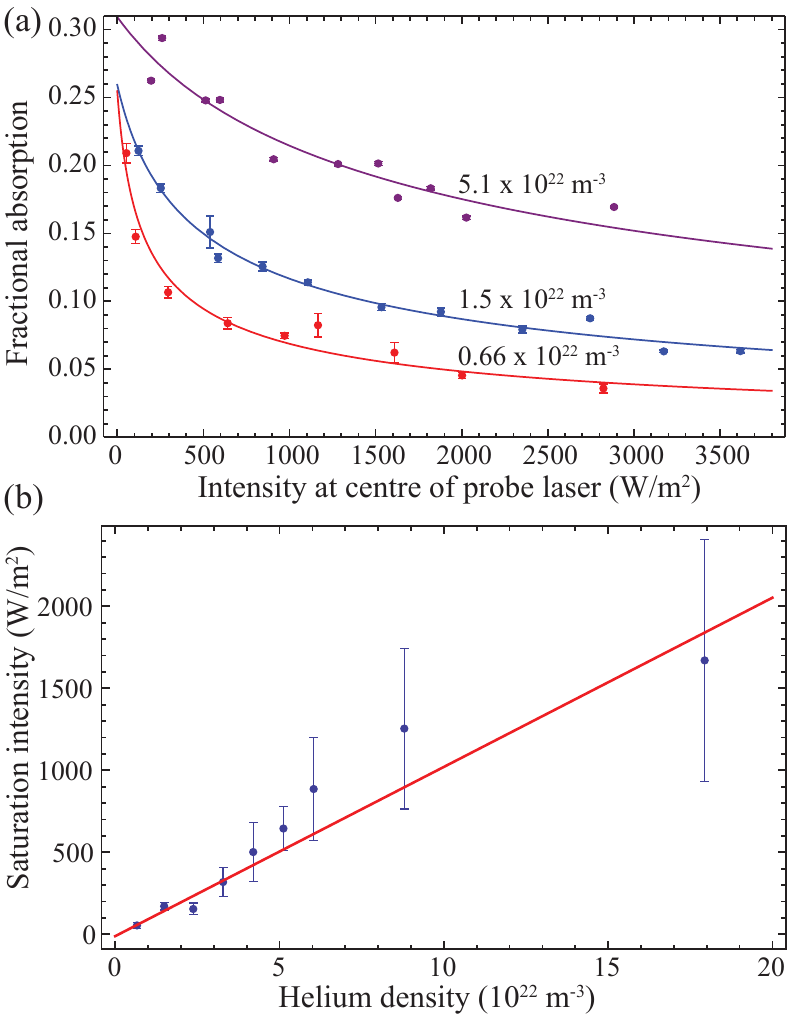}
\caption{\label{fig:saturation}(Color online)
(a) Fractional absorption versus probe intensity at three different helium densities. The temperature is 80\,K. (b) Saturation intensity versus helium density at 80\,K.
}
\end{figure}

Figure\,\ref{fig:saturation}(a) shows how the fractional absorption of the probe laser depends on the laser intensity for three different values of the buffer gas density. In these experiments, the temperature was 80\,K. As the laser intensity increases, the fractional absorption decreases. This fall in fractional absorption is less rapid when the helium density is high than when it is low. The data show that there is a saturation of the absorption, and that the intensity at which this saturation sets in increases with increasing buffer gas density. It is interesting to explore the mechanism responsible for this saturation.

In a two level system, the absorption saturates once the rate of excitation from ground to excited state exceeds the rate of decay of the excited state. Once this occurs the rate at which photons are absorbed is limited by the decay rate, and an increase in the incident power does not produce a proportional increase in the absorbed power. Collisions increase the decay rate, resulting in an increase in the intensity required to reach saturation, accompanied by broadening of the spectral line. Our saturation data show a strong dependence on the helium density, but we observed no pressure broadening of the spectral lines in our saturated absorption spectra for densities up to $20 \times 10^{22}$\,m$^{-3}$. We conclude that this is not the mechanism responsible for the saturation observed here.

The molecules are not well modelled by a two level system, but they can be fairly well modelled using three levels. The probe laser connects levels 1 and 2, while level 3 represents all other levels in the molecule and is not coupled at all to the laser. Since level 2 can decay to level 3, the molecules are optically pumped into level 3 by the probe laser. The intensity needed to saturate the absorption is set by a competition between the optical pumping of molecules out of level 1, and any process that puts population back into this level. We identify three such processes. Firstly, diffusion brings new molecules into the interaction region. Secondly, inelastic collisions transfer molecules from level 3 back to level 1. Thirdly, when the Doppler width is much larger than the natural linewidth, optical pumping into level 3 occurs only in the velocity group that is resonant with the laser. Elastic collisions then repopulate that velocity group from the reservoir of molecules with all other velocities. The rate for the first process is inversely proportional to the helium density and is very small compared to all other relevant rates over the entire range of densities used here. The rates for the second and third processes are proportional to the helium density and, together, are responsible for the saturation of the absorption that we observe.

The expected fractional absorption is calculated in Appendix \ref{App1}. To model our data, we first make the small absorption approximation which simplifies the results considerably. In this approximation we assume that the laser-driven excitation rate is constant throughout the length of the absorbing column of gas, because the change in the intensity as the laser propagates through the column is small. This allows us to replace the variable intensity $I$ with the constant value $I_{\text{in}}$ on the right hand side of Eq.\,(\ref{Eq:dsdzIntegrated}). The maximum fractional absorption is 30\% for the data analyzed here and shown in Fig.\,\ref{fig:saturation}. This gives a rough indication of the maximum error that we make by assuming a constant excitation rate throughout the sample. Next we make the assumption, which we will verify later, that all the relevant collision rates are considerably smaller than the natural decay rate of the upper level, $\Gamma$. In this case, the fractional absorption is given by Eq.\,(\ref{Eq:LowAbsorptionIntegrated}) with the saturation intensity $I_{s}$ replaced by $I_{s}'$ which is given by Eq.\,(\ref{Eq:Is2}). The ratio of the Doppler width to the natural linewidth, $w$, enters the expression for the fractional absorption, and its value given by Eq.\,(\ref{Eq:w}) is $w=15$ in this experiment. This being much larger than 1, the fractional absorption varies with intensity as $(1+I_{\text{in}}/I_{s}')^{-1/2}$. Equation \,(\ref{Eq:LowAbsorptionIntegrated}) assumes a constant intensity across the laser beam but the probe beam used in these experiments had a Gaussian intensity distribution. We need to account for this in our analysis by integrating over the distribution. The result for the fractional absorbed power, which is the quantity measured in the experiment, is then
\begin{multline}
\bar{{\cal A}}\!=\!\int_{0}^{\infty}\frac{\sqrt{\pi}\alpha_{0}\,e^{-\rho^{2}}\,e^{\frac{1}{4w^{2}}
\left(1\!+\!\frac{I_{0}}{I_{s}'}e^{-\rho^{2}}\right)}}{w \sqrt{1\!+\!\frac{I_{0}}{I_{s}'}e^{-\rho^{2}}}} \times \\
\left[1\!-\!\text{erf}\left( \frac{1}{2w}\sqrt{1\!+\!\frac{I_{0}}{I_{s}'}e^{-\rho^{2}}}\right)\right] \rho\,d\rho,
\label{Eq:FractionalPowerAbsorbed}
\end{multline}
where $I_{0}$ is the intensity at the centre of the probe beam. The quantity $I_{s}'$ is a modified saturation intensity and is given by Eq.\,(\ref{Eq:Is2}), while $\alpha_{0}$ is the optical thickness of the sample for laser light resonant with the $1 \rightarrow 2$ transition and is given by Eq.\,(\ref{Eq:alpha0}). We fit the data shown in Fig.\,\ref{fig:saturation}(a), along with similar datasets taken at other helium densities, to Eq.\,(\ref{Eq:FractionalPowerAbsorbed}) with $\alpha_{0}$ and $I_{s}'$ as the only free parameters in the fit. The solid lines in Fig.\,\ref{fig:saturation}(a) are the results of this fit for the three datasets shown there.

Figure \ref{fig:saturation}(b) shows how the value of $I_{s}'$ obtained from these fits depends on the helium density, along with a weighted linear fit to these data. The linear dependence of the saturation intensity on the helium density supports our assumption that the collision rates are small compared with $\Gamma$, at least for the lower density data which has the greatest weight in the fit. When the collision rates become comparable with $\Gamma$, we expect terms of higher order in the density to appear in the expression for the saturation intensity, as shown by Eq.\,(\ref{Eq:beta}). A weighted fit to quadratic and cubic polynomials shows the coefficients of these higher-order terms to be zero within the error of the fit. The intercept of the linear fit is also zero within the fit error, as we would expect, and the values of $\alpha_{0}$ are independent of helium density. We conclude that our model is the correct one. We see from Eq.\,(\ref{Eq:Is2}) that the slope of the straight line fit gives us the ratio $(\gamma_{13}+\gamma_{c})/n_{\text{He}}$, where $\gamma_{13}$ is the rate at which collisions transfer population from level 1 to level 3, and $\gamma_{c}$ is the rate for velocity-changing collisions. We do not know which of these two rates, $\gamma_{13}$ or $\gamma_{c}$, is the larger, and our data does not help us to distinguish between them, so we simply take them together. Both rates are responsible for re-filling the hole produced by the laser in the combined velocity and internal-state distribution. We define the thermally-averaged cross-sections, $\sigma_{13}$ and $\sigma_{c}$, via
\begin{equation}
\gamma_{X} = n \sigma_{X} \sqrt{\frac{8 k T}{M_{\text{He}}\pi}}.
\end{equation}
The sum of the two cross-sections is obtained from the gradient $g = 1.0 \times 10^{-20}$\,(W\,m$^{-2}$)/m$^{-3}$ of the straight line fit:
\begin{equation}
\sigma_{13} + \sigma_{c} = \frac{3 \lambda^{3} r (1-r) p g}{2 \pi h c}\sqrt{\frac{M_{\text{He}}\pi}{8 k T}}.
\label{Eq:sigmahf}
\end{equation}

Here, $p$ is a factor that depends on the number of magnetic sub-states in levels 1 and 2, and is given by Eq.\,(\ref{Eq:p}). For this experiment its value is $p=1/3$. The branching ratio $r$ is the probability that the excited state decays back to the state resonant with the laser. For our experiment, it is the product of the Franck-Condon factor, $Z$, for the $0-0$ vibrational transition, and an angular factor that accounts for the probability of returning to the same rotational and hyperfine state. The calculation of this angular factor is similar to the one presented in the appendix of reference \cite{Wall:2008}.

Consider driving a transition from a positive-parity ground state $X^{2}\Sigma^{+}(v,N,J=N\pm1/2)$ to a negative-parity excited state $A^{2}\Pi_{1/2}(v',J'=N+1/2)$, where the spin-orbit splitting of the excited state is large compared to the rotational splitting, there are no intermediate electronic levels, and we neglect for the moment the hyperfine structure arising from any nuclear spin. The only possible decay routes are to the states $X^{2}\Sigma^{+}(v'',N'',J'')$ where $v''$ can take any value with weight given by the Franck-Condon factor for $v'\rightarrow v''$, but $(N'',J'')$ are constrained to one of three possibilities, (1) $(N'' = N, J''=N + 1/2)$ (2) $(N'' = N, J''= N - 1/2)$ or (3) $(N'' = N + 2, J''=N + 3/2)$. The corresponding transitions are usually called Q$_{11}$, R$_{12}$, and $P_{12}$, respectively. We find that the relative branching ratios between these three depends on the value of $J'$. When $J'=1/2$ the three branches have relative weights of 2/3,0,1/3, and as $J'$ increases these tend to the asymptotic values 1/2, 1/4, 1/4. For the data analyzed here the excitation was to $J'=27/2$, and the ratios are 0.50, 0.24, 0.26, very close to their asymptotic values. In excitation, the two transitions with $J=N\pm1/2$, Q$_{11}$ and R$_{12}$, are unresolved in our Doppler-limited experiment,  but are separated by more than the natural linewidth. Both transitions are excited, but the velocity group is different for the two excitation routes. In this case, we should take the mean value of $r$ for the two branches. The situation is further complicated by the fact that each transition consists of a pair of hyperfine components which are unresolved in the Doppler-limited spectrum. In the case of the Q$_{11}$ line the components are resolved in the Doppler-free spectrum and we should take the two hyperfine states as distinct. This halves the hyperfine-averaged value of $r$ for this line. For the $R_{12}$ line, the situation is different because the hyperfine components are unresolved in the Doppler-free spectrum. In this case it is better to treat the two hyperfine levels as a single level and the value of $r$ is unchanged. These considerations lead us to an averaged branching ratio of $r\approx0.25 Z$. We do not know of any direct measurements or calculations of the Franck-Condon factor for the A$(v'=0)$-X$(v''=0)$ transition of YbF. However, we know that when the molecule is excited strongly on the $F=1$ component of the Q$_{11}(1/2)$ line, the mean number of photons scattered is $1.9\pm0.1$ \cite{Tarbutt:2002}. This information, together with the table of branching ratios given in \cite{Wall:2008} provides an estimated value of $Z=0.91\pm0.05$. A high value for $Z$ is to be expected since YbF closely resembles the alkaline-earth monofluorides which all have this property \cite{Pelegrini:2005}. Finally then, we take $r\approx 0.23$ for this experiment.

Putting these values into Eq.\,(\ref{Eq:sigmahf}) we obtain $\sigma_{13} + \sigma_{c} = 37\times 10^{-20}$\,m$^{2}$. There are a number of uncertainties in determining this value. The fractional error on the gradient of the straight line fit is 15\%. Using the spread in the diffusion lifetimes measured at the same nominal helium density we estimate the fractional uncertainty in our measurement of the helium density to be 20\%. The fractional uncertainty in calibrating the laser intensity is also 20\%. The fractional uncertainty in the branching ratio $r$ is estimated to be 24\%, partly from the 5\% uncertainty in the Franck-Condon factor, but mostly from the uncertainty of how best to treat the four unresolved components of the line as discussed above. This leads to a 19\% error in the cross-section which depends on the value of $r(1-r)$. Other possible errors come from the approximations we have made in our model. First, we have assumed that the absorption is small; this does not contribute a significant error at this level. Second, our analysis relies on the approximation that the collision rates are much smaller than $\Gamma$. We have previously estimated $\Gamma$ to be approximately $6 \times 10^{7}$\,rad\,s$^{-1}$ \cite{Skoff:2009}. When we use our cross-section to calculate the collision rate, we obtain $\gamma_{13}+\gamma_{c}=5 \times 10^{7}$\,s$^{-1}$ at $n=2 \times 10^{23}$\,m$^{-3}$. This implies that the collision rate is approaching $\Gamma$ at the upper end of the plot. However, it is the points at low helium densities that primarily fix the gradient of the fit, since these have far more weight in the fit, and for these values the collision rates are much smaller than $\Gamma$, consistent with our original assumption. Third, we have used a fitting function that is appropriate when the laser is resonant for molecules at rest, but this cannot be satisfied for all four of the unresolved components. We find that setting $\delta_{L}'= w/2$ decreases the cross-section by 3\%, and setting $\delta_{L}'= w$ decreases the cross-section by 12\%. Considering the mean displacement of the components from the line centre at 80\,K, we conclude that the resulting change in the cross-section, and the associated error, are both negligible for this measurement. The final result is $\sigma_{13} + \sigma_{c} = 37 \pm 14 \times 10^{-20}$\,m$^{2}$. This result is close to our calculated YbF-He diffusion cross-section at 80\,K, given in Table \ref{Tab:cross-sections}, as one would expect if velocity-changing collisions dominate over inelastic collisions in determining the saturation intensity.

The values of $\alpha_{0}$ obtained from the fits to the data shown in Fig.\,\ref{fig:saturation}(a), along with similar datasets taken at other helium densities, are independent of helium density and are consistent with one another across the whole range of helium densities. The mean value is $\alpha_{0}=18.1 \pm 0.8$. The fact that $\delta_{L}\ne 0$ for the four unresolved components has a much larger effect on $\alpha_{0}$ than on $I_{s}'$, and by considering the effect of this we obtain a corrected value of $\alpha_{0}=f_{1}n\sigma L=23\pm4$. Using the above values of $r$ and $p$, the value for the optical absorption cross-section given by Eq.\,(\ref{Eq:CrossSectionSimple}) is $\sigma=1.1 \times 10^{-14}$\,m$^{2}$. Taking an absorption length of $L=3$\,cm and the fraction of all molecules in $N=13$ to be $f_{1}=0.05$ at 80\,K, we find the total density of YbF molecules to be $n \approx 1.3 \times 10^{18}$\,m$^{-3}$ in these experiments. The total number of YbF molecules produced is approximately $3 \times 10^{13}$. Of course, the YbF density depends on the energy and spot size of the ablation laser, the target quality and the buffer gas density. Nevertheless, we consistently produce this density of molecules once the experimental conditions are optimized.

\section{Conclusions}

We have investigated the dynamics of YbF molecules produced inside a helium buffer gas cell using a combination of absorption imaging, absorption spectroscopy and saturated absorption spectroscopy. While we have focussed on this particular molecule, we expect our results to carry over to a wide range of other polar radicals produced using similar methods.

The absorption images show that, upon ablation of the target, the molecules that are formed expand ballistically into the cell on a $\mu$s timescale. When the helium density is low, this initial ballistic expansion completely fills the cell, but at higher helium density the expansion is arrested with the molecules confined to a region around the target. The molecules then diffuse slowly to the walls of the cell. We have measured how the diffusion time depends on the helium density and find the expected linear dependence at low density. At higher densities the diffusion times no longer increase with density but instead reach a roughly constant value. Our modelling of the experiment shows that the diffusion is more rapid when the molecules are localized near the target than when they fill the cell, and that in the former case the decay curves depend strongly on the probe position, with the density decaying much faster when the probe is close to the target. This observation, along with the insights obtained from the absorption imaging, suggest that the levelling off of the lifetimes is due, at least in part, to the initial confinement of the molecules to a region close to the target. By comparing the diffusion of YbF and Li at low helium densities, and using values for the Li diffusion cross-section calculated from a well-established interaction potential, we have measured the YbF-He diffusion cross-section at two temperatures. The experimental values agree with the theoretical values we calculate from the YbF-He potential given in \cite{Tscherbul:2007}. This is a first experimental test of that potential. We have compared our experimental results for the Li diffusion times with the results of a numerical model of diffusion in this cell. The results agree at 80\,K and 20\,K, but differ by a factor of two at 293\,K where the results are most precise. This discrepancy is not understood but suggests that diffusion cross-sections determined by measuring lifetimes in a cell may not be accurate unless a reference atom is used.

We have measured how the translational and rotational temperatures of the molecules change as a function of the time since their formation, for typical helium densities of about $10^{22}$\,m$^{-3}$, and have compared our experimental results with those of various thermalization models. Our conclusion is that the YbF molecules thermalize with the helium more rapidly than the $50$\,$\mu$s time resolution of the experiment. Instead of observing that thermalization process, we see the cooling of the helium buffer gas which heats up when the target is ablated and then slowly cools back down to the temperature of the cell walls. We see that the cooling occurs on two separate timescales which differ by a factor of 10, and we suppose that the heat of ablation is deposited in a local region close to the target, and that this heat first dissipates quickly into the rest of the gas, and then more slowly as it diffuses to the walls.

Finally, we have made a detailed study of the absorption of a probe laser by molecules in a buffer gas cell, elucidating the main dynamical processes that are important. As the power of the probe laser is increased the absorbed power increases at first, but eventually saturates because the population in the resonant state is optically pumped into a multitude of other states that are not resonant with the laser. The resulting hole in the combined velocity and internal state distribution is repopulated by collisions, and so saturation of the absorption sets in when the rate of optical pumping is equal to the relevant collision rates. We have measured the fractional absorption as a function of laser intensity over a range of helium densities and find that the data fits well to our model. The intensity needed to saturate the absorption is directly proportional to the helium density for densities up to $10^{22}$\,m$^{-3}$. The cross-section for the ``hole-filling'' collisions is consistent with our theoretical value for the YbF-He diffusion cross-section, as we would expect if velocity-change is the dominant hole filling mechanism. The same data are used to estimate the total YbF density, which is approximately $10^{18}$\,m$^{-3}$. The total number of YbF molecules produced is approximately $3 \times 10^{13}$.

Using an improved setup, we have now reduced the cell temperature to 3.5\,K. At this temperature there will be about $3 \times 10^{12}$ molecules in the rotational ground state. Using a hydrodynamic flow of helium to sweep the molecules out of the cell, we expect that at least 10\% can be extracted into a beam with a divergence of about 0.1\,steradian \cite{Patterson:2007}, resulting in a cold, slow-moving molecular beam with a flux of $3 \times 10^{12}$ ground state YbF molecules per steradian per pulse. This is 2000 times more intense than the flux produced in a supersonic source of the same molecules \cite{Tarbutt:2002}, and provides molecules at considerably lower speed, and so promises a large improvement in the sensitivity of molecular beam experiments such as the measurement of the electron's electric dipole moment.

\acknowledgements
We are grateful to Jon Dyne, Steve Maine and Valerijus Gerulis for their expert technical assistance. We thank Roman Krems for sending us the YbF-He interaction potential. This work was supported by the EPSRC, the STFC, and the Royal Society. The research leading to these results has received funding from the European Community's Seventh Framework Programme FP7/2007-2013 under the grant agreement 216774.

\appendix
\section{Laser absorption by molecules in the presence of a buffer gas}
\label{App1}

To understand the interaction of the probe laser with the molecules in the buffer gas, we use a 3 level model of the molecule. In this model, a plane polarized laser field, ${\bf E}=E_{0}{\bf\hat{z}}\cos(\omega t)$, is detuned from resonance with the transition from level 1 to level 2 by an angular frequency $\delta$. Level 3 represents all other relevant molecular levels and is not coupled to the laser field. Level 2 can decay by spontaneous emission, either to level 1 at rate $r \Gamma$, or to level 3 at rate $(1-r) \Gamma$. Inelastic collisions with helium atoms also cause level 2 molecules to decay to levels 1 and 3, with rates $\gamma_{21}$ and $\gamma_{23}$ respectively. Elastic collisions dephase level 2 at a rate $\gamma_{el}$. Levels 1 and 3 are assumed to be in the ground electronic state of the molecule, so spontaneous emission out of these levels is negligible. Collisions result in population transfer from level 1 to level 3, and vice versa, with rates $\gamma_{13}$ and $\gamma_{31}$ respectively. There is not enough energy for a collision to excite molecules to level 2. Molecules move in and out of the interaction region with a characteristic rate $\phi$, which is the inverse of the average time taken to diffuse out of the probe laser beam. Following the usual treatment that leads to the optical Bloch equations for a two level atom, modified to account for the third level, for the collisions, and for the flux, we obtain the following equations for the rate of change of the level 1 and 2 populations, $\rho_{11}$ and $\rho_{22}$, and of the coherence between these two levels $\rho_{12}$:

\begin{gather}
\dot{\rho}_{11} = i\frac{\Omega }{2}\left(\rho _{12}-\rho _{21}\right)+\left(r \Gamma +\gamma _{21}\right) \rho _{22}-\gamma _{13}\rho _{11} \nonumber \\
+\gamma _{31}\left(1-\rho _{11}-\rho _{22}\right)-\phi  \rho _{11}+f_1 \phi,\label{Eq:rho11}\\
\dot{\rho}_{22} = -i\frac{\Omega }{2}\left(\rho _{12}-\rho _{21}\right)-\left(\Gamma +\gamma _{21}+\gamma _{23}\right) \rho _{22}-\phi  \rho _{22},\\
\dot{\rho}_{12}\!=\!i\frac{\Omega }{2}\left(\rho _{11}\!-\!\rho _{22}\right)\!+\!\left(i \delta\!-\!\frac{1}{2}\left(\Gamma\!+\!\gamma _{21}\!+\!\gamma _{23}\right)\!-\!\gamma _{\text{el}}\right) \rho _{12}.
\end{gather}
Here, $f_{1}$ is the fraction of all molecules in level 1 when the probe laser is off, and $\Omega$ is the Rabi frequency defined by $\Omega = E_{0} z_{12}/\hbar$ where $z_{12}$ is the $z$ component of $d_{12}=\langle 1 | {\bf d} |2\rangle$ and ${\bf d}$ is the dipole moment operator. As the laser propagates a small distance $\Delta z$ through the molecular sample of number density $n$, its intensity, $I=c \epsilon_{0} E_{0}^{2}/2$, changes due to absorption and stimulated emission by the amount

\begin{equation}
\Delta I = i \hbar  \omega  \frac{\Omega }{2} \left(\rho _{12}-\rho _{21}\right)n \Delta z.
\end{equation}

It will be useful to define the dimensionless rates $\gamma_{ij}'=\gamma_{ij}/\Gamma$, $\gamma_{el}'=\gamma_{el}/\Gamma$, $\delta'=\delta/\Gamma$, $\phi'=\phi/\Gamma$ and $\Gamma_{T}'=1+\gamma_{21}'+\gamma_{23}'+2 \gamma_{el}'$, along with a dimensionless quantity proportional to the laser intensity, $s=2\Omega^{2}/\Gamma^{2}=I/I_s$, where $I_s$ is called the saturation intensity. It will also be helpful to make use of the relationship between $\gamma_{31}$ and $\gamma_{13}$ which is easily found by ensuring that the populations are in a steady state when the laser is turned off:
\begin{equation}
\frac{\gamma_{13}}{\gamma_{31}} = \frac{1-f_{1}}{f_{1}}.
\end{equation}

We will calculate the absorption averaged over a time window that is very short compared to the characteristic diffusion time, but very long compared to $1/\Gamma$. In this case, we can take $n$ to be a constant, and we can use the steady state solution to the above dynamical equations. After some algebra we find the steady state values for the density of molecules in levels 1 and 2 to be
\begin{gather}
n_{1,ss}=n_{2,ss}\left(1+\frac{\left(\Gamma +\gamma _{21}+\gamma _{23}+\phi \right)}{R}\right), \label{Eq:SteadyState1}\\
n_{2,ss} = \frac{R\left(\gamma _{31}+f_{1} \phi \right)n}{X + Y},\label{Eq:SteadyState2}\\
X = \left(\gamma _{13}+\gamma _{31}+\phi \right)\left(\Gamma +\gamma _{21}+\gamma _{23}+\phi \right),\\
Y = R\left(\gamma _{13}+\gamma _{23}+2\gamma _{31}+\Gamma (1-r)+2\phi \right),
\end{gather}
where $R$ is the laser excitation rate,
\begin{equation}
R=\frac{\Gamma }{2}\left(\frac{\Gamma_{T}'}{4\delta'^{2}+\Gamma_{T}'^{2}}\right)s.
\label{Eq:LaserRate}
\end{equation}

With the populations in steady-state, the equation for the intensity reduces to
\begin{equation}
\frac{d I}{d z} = -\hbar \omega (\Gamma + \gamma_{21} + \gamma_{23} + \phi)n_{2,ss}.
\end{equation}
It is convenient to rewrite this equation in terms of the reduced intensity $s=2\Omega^{2}/\Gamma^{2}$, which we can do using the relationships between $I$, $E_{0}$, and $\Omega$ given above. We also introduce a reduced length $Z=z/L$, $L$ being the total length of the interaction region. Thus we obtain
\begin{equation}
\frac{d s}{d Z}=-2\sigma L(1+\gamma_{21}'+\gamma_{23}'+\phi')n_{2,ss},
\label{Eq:dsdz}
\end{equation}
where
\begin{equation}
\sigma=\frac{3\lambda^{2}}{2\pi} \frac{z_{ij}^{2}}{\sum_{k} d_{jk}^{2}}
\label{Eq:CrossSection}
\end{equation}
is the absorption cross-section of the molecule. To reach Eq.\,(\ref{Eq:CrossSection}) we have used the expression for the decay rate, $\Gamma =\frac{1}{3\pi  \epsilon _0\hbar  c^3}\sum _k\omega_{jk}^3 d_{jk}^2$, have made the approximation that the frequencies of all allowed transitions are equal, $\omega_{jk}=2\pi c/\lambda$, and have generalized the notation in order to handle the prickly issue of degeneracy. To represent the molecule accurately, we have to allow each of levels 1 and 2 to represent a set of degenerate or near-degenerate levels. In Eq.\,(\ref{Eq:CrossSection}), $i$ represents a particular Zeeman sub-level of level 1, $i=(1,M_{p})$, $j$ is the sub-level $(2,M_{p})$ of level 2, and $k$ is an index which runs over all the sub-levels of levels 1 and 3. The branching ratio $r$ is the rate for spontaneous decay to {\it any} of the sub-levels in 1, divided by the total spontaneous decay rate. Using this, we see that Eq.\,(\ref{Eq:CrossSection}) can be written in the form
\begin{equation}
\sigma=\frac{3 \lambda ^2 r p}{ 2\pi },
\label{Eq:CrossSectionSimple}
\end{equation}
where
\begin{equation}
p = \frac{|\langle 1,M_p|\pmb{d}\pmb{.}\hat{\pmb{z}}|2,M_p\rangle |^2}{\sum _{M_1}|\langle 2,M_p|d|1,M_1\rangle |^2}
\label{Eq:p}
\end{equation}
is a factor that accounts for the degeneracy, the sum running over all sublevels $M_{1}$ of level 1. Finally, when level 1 is degenerate and the molecular sample is unpolarized, we have to remember to take the average value of $\sigma$ calculated for each of the ground state sub-levels $M_p$. Using these same definitions, we can write the saturation intensity as
\begin{equation}
I_{s} = \frac{\pi  h c \Gamma}{3\lambda ^3 r p}.
\label{Eq:Is}
\end{equation}

Returning to the calculation of the absorption we use Eq.\,(\ref{Eq:SteadyState2}) to substitute for $n_{2,ss}$ in Eq.\,(\ref{Eq:dsdz}), leading us to the result
\begin{equation}
\frac{d s}{d Z} = -\frac{\alpha s}{1+\beta s},
\label{Eq:dsdzFinalForm}\\
\end{equation}
where
\begin{equation}
\alpha = \frac{f_{1} n \sigma L}{\Gamma_{T}'} \left(\frac{\Gamma_{T}'^{2}}{\Gamma_{T}'^{2} + 4 \delta'^{2}}\right) =\alpha_{0}\left(\frac{\Gamma_{T}'^{2}}{\Gamma_{T}'^{2} + 4 \delta'^{2}}\right),
\label{Eq:alpha}
\end{equation}
and
\begin{align}
\beta &\!=\! \frac{1-r+\gamma_{13}'+\gamma_{23}'+ 2\gamma_{31}'+2\phi'}{2\Gamma_{T}'(\gamma_{13}'+\gamma_{31}'+\phi')(1+\gamma_{21}'+\gamma_{23}'+\phi')}\left(\frac{\Gamma_{T}'^{2}}{
\Gamma_{T}'^{2}+4\delta'^{2}}\right)\nonumber\\
&= \beta_{0}\left(\frac{\Gamma_{T}'^{2}}{\Gamma_{T}'^{2}+4\delta'^{2}}\right).
\label{Eq:beta}
\end{align}
These equations define $\alpha_{0}$ and $\beta_{0}$ which are the values of $\alpha$ and $\beta$ when $\delta=0$.

Integrating Eq.\,(\ref{Eq:dsdzFinalForm}) we obtain the result
\begin{equation}
\ln\left(\frac{s}{s_{\text{in}}}\right) + \beta(s- s_{\text{in}}) = -\alpha Z,
\label{Eq:Log}
\end{equation}
where $s_{\text{in}}$ is the value of the reduced intensity at the entrance of the cell. We have taken $n$ (and hence $\alpha$) to be independent of $Z$ over the interaction length. If it is not, we have simply to replace the product $n Z$ by the column density $\int n(Z)dZ$. We can rewrite this transcendental equation in the form
\begin{equation}
\beta s \, e^{\beta s}=\beta s_{\text{in}}\, e^{-\alpha  Z+ \beta s_{\text{in}}},
\end{equation}
which has the solution
\begin{equation}
s(Z) = \frac{1}{\beta}\, W\left(\beta s_{\text{in}}\, e^{(-\alpha  Z+ \beta s_{\text{in}})}\right).
\end{equation}
Here $W(x)$  is the Lambert W function\footnote{The Lambert W function is sometimes known as the omega function or the product log. It is the inverse function of $f(W) = W e^{W}$}. The fractional absorption measured at the end of the interaction region (i.e. at $Z=1$ ) is
\begin{equation}
{\cal A} = \frac{s_{\text{in}} - s_{\text{out}}}{s_{\text{in}}} = 1 - \frac{1}{\beta s_{\text{in}}} W\left(\beta s_{\text{in}}\,e^{(-\alpha + \beta s_{\text{in}})}\right).
\label{Eq:FracAbsorptionComplete}
\end{equation}

So far we have neglected the distribution of speeds in the gas. Therefore, our results are only valid in the case where the Doppler width is small compared to $\Gamma$. This is not often the case, and so we need to extend the treatment to include the distribution of speeds, $g(v)$, where $v$ is the component of velocity along the propagation direction of the probe laser. This modifies the treatment in two ways. First, we have to write the equations for the absorption as a function of $v$ and then integrate over $v$ at the end to obtain the total absorption. Let us use a new symbol, $\delta_{L}$, for the detuning of the laser from the at-rest transition frequency, and use $\delta_{D}$ for the Doppler shift of a molecule, $\delta_{D} = 2\pi v/\lambda$. Then, $\delta=\delta_{L}-\delta_{D}$. Instead of Eq.\,(\ref{Eq:dsdzFinalForm}), we then have
\begin{equation}
\frac{d s}{d Z} = -\int \frac{\alpha(v) g(v) s}{1+\beta(v) s} dv,
\label{Eq:dsdzIntegrated}\\
\end{equation}
where $\alpha(v)$ and $\beta(v)$ are the same as given by Eq.\,(\ref{Eq:alpha}) and Eq.\,(\ref{Eq:beta}) but with $\delta'$ replaced by $\delta_{L}'-\delta_{D}'$. All the other symbols have the same meaning as before. The velocity distribution is
\begin{align}
g(v)\,dv &= \sqrt{\frac{M}{2\pi k T}}\exp\left(\frac{-M v^{2}}{2 k T}\right)\,dv \nonumber \\
&= \frac{1}{\sqrt{\pi}w} \exp\left(-\frac{\delta_{D}'^{2}}{w^{2}}\right)\,d\delta_{D}'.
\label{Eq:speedDist}
\end{align}
Here, the width $w$ is related to the temperature $T$ and the molecular mass $M$ through
\begin{equation}
w^{2}=\frac{8\pi^{2} k T}{M \lambda^{2}\Gamma^{2}}.
\label{Eq:w}
\end{equation}

The second modification that we need to make is to account for velocity-changing collisions in the rate equations. The probe laser tends to modify the velocity distribution by depleting the resonant velocity group, whereas collisions tend to restore the velocity distribution towards thermal equilibrium. The extra terms that appear in the rate equation for $n_{1}$ are
\begin{equation}
\left[\dot{n}_1(v)\right]_c=-n_1(v)\int \gamma _c(v',v)dv'+\int \gamma _c(v,v')n_1(v')dv',
\end{equation}
where $\gamma _c(v',v)dv'$ is the rate for collisions that change the velocity from $v$ to a value in the range $v'\rightarrow v'+dv'$. We make the ``strong collisions'' approximation, meaning that $\gamma_c(v',v)$ is independent of $v$ \cite{Smith:1971}. Then, to make sure that the system is in equilibrium when the probe laser is turned off, we must have
\begin{equation}
-n_1(v)\int \gamma _c(v')dv'+\gamma _c(v)\int n_1(v')dv'=0,
\end{equation}
from which it follows that $\gamma_c(v)=g(v)\gamma_{c}$, where $\gamma_{c}$ is the total collision rate integrated over all final velocities, $\gamma_{c}=\int \gamma _c(v')dv'$. It follows that the extra terms that appear in the rate equation for $n_{1}$ reduce to
\begin{equation}
\left[\dot{n}_1(v)\right]_c=-n_1(v)\gamma _c+f_1 g(v)\gamma _c n.
\end{equation}

If we look back at Eq.\,(\ref{Eq:rho11}), re-write it as a function of $v$ and multiply through by $n$ so that it becomes a rate equation for $n_{1}(v)$, we find that the last two terms in the equation are $-n_{1}(v)\phi + f_{1} g(v) \phi n$, which has exactly the same form as the above equation. A similar replacement occurs in the equation for $\rho_{22}$. So we see that to include velocity-changing collisions we simply have to replace $\phi$ with $\phi + \gamma_{c}$.

To summarize, the change in the probe intensity is given by Eq.\,(\ref{Eq:dsdzIntegrated}) where $g(v)$ is given by Eq.\,(\ref{Eq:speedDist}), $\alpha(v) = \alpha_{0} \Gamma_{T}'^{2}/(\Gamma_{T}'^{2} + 4 \delta'(v)^{2})$, $\beta(v) = \beta_{0}\Gamma_{T}'^{2}/(\Gamma_{T}'^{2}+4\delta'(v)^{2})$ and $\alpha_{0}$ and $\beta_{0}$ are
\begin{gather}
\alpha_{0} = \frac{f_{1} n \sigma L}{\Gamma_{T}'} \label{Eq:alpha0}\\
\beta_{0} = \frac{1-r+\gamma_{13}'+\gamma_{23}'+ 2\gamma_{31}'+ 2\gamma_{c}' + 2\phi'}{2\Gamma_{T}'(\gamma_{13}'+\gamma_{31}'+\gamma_{c}' + \phi')(1+\gamma_{21}'+\gamma_{23}'+\gamma_{c}' + \phi')}.
\end{gather}

These are very general results that can be used to calculate the degree of absorption for almost any case, once the relevant rates are known. The same equations give the spectral lineshapes to be expected as the laser frequency is scanned, and tell us how the fractional absorption changes with the intensity of the probe laser. In general, we will have to solve Eq.\,(\ref{Eq:dsdzIntegrated}) numerically. Furthermore, if the intensity of the probe laser is not uniform over the interaction volume, we should divide up the volume into small elements and integrate over these too. However, in many cases of interest the results simplify considerably because some particular quantity is small. Next, we examine some of these simpler cases.

\subsection{The low intensity limit}

When $\beta s_{\text{in}} \ll 1$ we can neglect it in the denominator of Eq.\,(\ref{Eq:dsdzIntegrated}), so that we have
\begin{equation}
\frac{d s}{d Z} = -\int \alpha(v) g(v) s dv = -\bar{\alpha} s
\end{equation}
where
\begin{equation}
\bar{\alpha}=\int \alpha(v) g(v)\,dv.
\end{equation}
The fractional absorption is
\begin{equation}
{\cal A} =  1 - s_{\text{out}}/s_{\text{in}} = 1 - e^{-\bar{\alpha}}.
\label{Eq:lowIntensityA}
\end{equation}
The full expression for $\bar{\alpha}$ is quite complicated, but when the laser is on resonance with the atomic transition (i.e. $\delta = \delta_{D}$) it simplifies to a reasonably compact form:
\begin{equation}
\bar{\alpha }=\alpha _0\frac{\sqrt{\pi}\,\Gamma_{T}'}{2w}e^{\Gamma_{T}'^{2}/\left(4w^2\right)}[1-\text{erf}(\Gamma_T'/(2w))].
\end{equation}
When $w \gg \Gamma_{T}'$, which will usually be the case if the gas is warm and the collision rates are not too much larger than $\Gamma$, this expression simplifies further to $\bar{\alpha }=\alpha _0 \sqrt{\pi}\,\Gamma_{T}'/(2w)$.

\subsection{The low absorption limit}

When the absorption is small the value of $s$ hardly changes and it is valid to set $s=s_{\text{in}}$ in the integral of Eq.\,(\ref{Eq:dsdzIntegrated}). Then, the fractional absorption is
\begin{equation}
{\cal A} = 1 - s_{\text{out}}/s_{\text{in}} = \int \frac{\alpha(v) g(v)}{1+\beta(v) s_{\text{in}}}\,dv.
\end{equation}
When the laser is on resonance, this integral evaluates to an analytical form,
\begin{equation}
{\cal A} = \frac{\alpha_{0}\sqrt{\pi}\,\Gamma_{T}'e^{\frac{\Gamma_{T}'^{2}}{4w^{2}}\left(1+\beta_{0} s_{\text{in}}\right)}}{2w \sqrt{1+\beta_{0} s_{\text{in}}}}\left[1-\text{erf}\left(\frac{\Gamma_{T}'}{2w}\sqrt{1+ \beta_{0} s_{\text{in}}}\right)\right].
\end{equation}
When the temperature is high so that $w \gg \Gamma_{T}'$, and the intensity is not too high, the expression simplifies to ${\cal A}=\alpha _0 \sqrt{\pi}\,\Gamma_{T}'/(2w \sqrt{1+\beta_{0}s_{\text{in}}})$.

\subsection{The two level limit with no collisions and no flux}

To verify that we can recover some familiar results, let us specialize to the case of just two levels by setting all the rates involving level 3 to zero and by setting $r=1$. Neglecting all the remaining collision rates and the flux term $\phi$, we obtain
\begin{gather}
\alpha = \frac{f_{1} n \sigma L}{1+4\delta'^{2}},\nonumber\\
\beta = \frac{1}{1+4\delta'^{2}}.\nonumber
\end{gather}
When the intensity is low and the velocity distribution is neglected, we have, using Eq.\,(\ref{Eq:lowIntensityA})
\begin{equation}
{\cal A} = 1 - \exp\left(\frac{-f_{1} n \sigma L}{1+4\delta'^{2}}\right).
\label{Eq:ExponentialAbsorption}
\end{equation}
When the absorption is small and the velocity distribution is neglected, but the input intensity is not necessarily small, we have
\begin{equation}
{\cal A} = \frac{f_{1} n \sigma L}{1+4\delta'^{2}+I_{\text{in}}/I_{s}},
\label{Eq:LowAbsorptionGeneral}
\end{equation}
where $I_s$ is the saturation intensity defined by Eq.\,(\ref{Eq:Is}). In Eqs.\,(\ref{Eq:ExponentialAbsorption}) and (\ref{Eq:LowAbsorptionGeneral}) we have kept $f_{1}$ in the numerator even though it is unity in this case because, as we shall soon see, the same equations hold for other cases. We note that in this two level limit, the absorption cross-section may reach its maximum possible value, $\sigma_{\text{max}}=3\lambda^{2}/(2\pi)$. For example, in an atomic transition with zero angular momentum in the ground state and one unit in the excited state (an s-p transition) we have $p=1$ and so $\sigma=\sigma_{\text{max}}$. By contrast a p-s transition has $p=1/3$ for the $M=0$ component and $p=0$ for the other two components meaning that for an unpolarized sample the level-averaged absorption cross-section is $\sigma = \sigma_{\text{max}}/9$.

When we include the velocity distribution and put the laser on resonance, we have
\begin{equation}
\bar{{\cal A}} = \frac{\sqrt{\pi}}{2w}\frac{f_{1} n \sigma L}{\sqrt{1+\frac{I_{\text{in}}}{I_{s}}}}e^{\frac{1}{4w^{2}}\left(1+\frac{I_{\text{in}}}{I_{s}}\right)}\left[1-\text{erf}\left(\frac{1}{2w}\sqrt{1+ \frac{I_{\text{in}}}{I_{s}}}\right)\right].
\label{Eq:LowAbsorptionIntegrated}
\end{equation}

\subsection{The low pressure, very low flux limit}

Returning to the 3 level system, let us assume that all the collision rates are much smaller than $ (1-r)\Gamma$, and let us also take the flux $\phi$ to be much smaller than the collision rates. This limit applies to most of the experiments reported in this paper.

In almost all situations we expect $f_{1} \ll 1$ because the effective degeneracy of state 3 is vastly larger than that of level 1. Then we have $\gamma_{13} \gg \gamma_{31}$. The only exception is when level 1 is the ground rotational state and the rotational temperature is much lower than the rotational spacing. Using this approximation, along with the ones discussed above, $\alpha$ and $\beta$ simplify to
\begin{gather}
\alpha_{0} \approx f_{1} n \sigma L,\nonumber\\
\beta_{0} \approx \frac{1-r}{2(\gamma_{13}'+\gamma_{c}')}.\nonumber
\end{gather}
When the laser intensity is low and we neglect the distribution of velocities the fractional absorption is again given by Eq.\,(\ref{Eq:ExponentialAbsorption}), just as in the two-level case. Similarly, when the absorption is small and we neglect the velocity distribution, we again have Eq.\,(\ref{Eq:LowAbsorptionGeneral}) except that $I_s$ in this equation is replaced by a new saturation intensity
\begin{equation}
I_{s}'=\frac{2}{1-r}\frac{\gamma_{13}+\gamma_{c}}{\Gamma}I_s.
\label{Eq:Is2}
\end{equation}
When the absorption is small and we include the velocity distribution, the result is the same as in Eq.\,(\ref{Eq:LowAbsorptionIntegrated}), but with $I_s$ again replaced by the $I_{s}'$ of Eq.\,(\ref{Eq:Is2}).

We see that for the three level system in this limit, the intensity required to saturate the absorption is proportional to the sum of two rates, the collision rate that transfers population between levels 1 and 3, and the rate for velocity changing collisions. Equations (\ref{Eq:LaserRate}) and (\ref{Eq:Is2}) together show that, when $\delta =0$ and $I=I_{s}'$, we have $(1-r)R = \gamma_{13} + \gamma_{c}$ (in this low pressure limit where $\Gamma_{T}'=1$). Since $1/(1-r)$ is the mean number of photons scattered by the molecule before being pumped into level 3, $(1-r) R$ is the optical pumping rate. So $I_{s}'$ is the intensity required to make the optical pumping rate equal to the rate at which collisions redistribute population between levels 1 and 3 or between one velocity group and another.

\subsection{The low flux, very low pressure limit}

In this limit we take $\phi \ll (1-r)\Gamma$ and take all the collision rates to be far smaller than $\phi$. In this case we have
\begin{gather}
\alpha_{0} \approx f_{1} n \sigma L,\nonumber\\
\beta_{0} \approx \frac{1-r}{2\phi'}.\nonumber
\end{gather}
Once again, the fractional absorption in both the low intensity limit and the low absorption limit is the same as in the two level case, but with $I_s$ now replaced by
\begin{equation}
I_{s}''=\frac{2}{1-r}\frac{\phi}{\Gamma}I_s.
\label{Eq:Is3}
\end{equation}
The intensity required to saturate the absorption is proportional to $\phi$. When $\delta =0$ and $I=I_{s}''$, we have $(1-r)R = \phi$, showing that $I_{s}''$ is the intensity for which the optical pumping rate equals the rate at which new molecules enter the interaction region.

\subsection{Summary}

When the intensity is low, the limiting rate is the laser excitation rate and the fractional absorption is ${\cal A} = 1 - e^{-\bar{\alpha}}$. The result does not depend on flux, velocity-changing collisions, rotation-changing collisions or vibration-changing collisions. Collisions only enter through the modification of the decay rate of level 2 from $\Gamma$ to $\Gamma_{T}$. When the intensity is not low the result is more complicated, but it simplifies if the degree of absorption is low. In that case the fractional absorption on resonance has the form given by Eq.\,(\ref{Eq:LowAbsorptionIntegrated}) where only the expression for the intensity needed to saturate the absorption changes from one limiting case to another. This saturation intensity is proportional to $\Gamma$ in the two-level limit, to $\gamma_{13}+\gamma_{c}$ in the low pressure, very low flux limit, and to $\phi$ in the low flux, very low pressure limit.


\begin{thebibliography}{46}
\expandafter\ifx\csname natexlab\endcsname\relax\def\natexlab#1{#1}\fi
\expandafter\ifx\csname bibnamefont\endcsname\relax
  \def\bibnamefont#1{#1}\fi
\expandafter\ifx\csname bibfnamefont\endcsname\relax
  \def\bibfnamefont#1{#1}\fi
\expandafter\ifx\csname citenamefont\endcsname\relax
  \def\citenamefont#1{#1}\fi
\expandafter\ifx\csname url\endcsname\relax
  \def\url#1{\texttt{#1}}\fi
\expandafter\ifx\csname urlprefix\endcsname\relax\def\urlprefix{URL }\fi
\providecommand{\bibinfo}[2]{#2}
\providecommand{\eprint}[2][]{\url{#2}}

\bibitem[{\citenamefont{Carr et~al.}(2009)\citenamefont{Carr, DeMille, Krems,
  and Ye}}]{Carr:2009}
\bibinfo{author}{\bibfnamefont{L.~D.} \bibnamefont{Carr}},
  \bibinfo{author}{\bibfnamefont{D.}~\bibnamefont{DeMille}},
  \bibinfo{author}{\bibfnamefont{R.~V.} \bibnamefont{Krems}}, \bibnamefont{and}
  \bibinfo{author}{\bibfnamefont{J.}~\bibnamefont{Ye}}, \bibinfo{journal}{New.
  J. Phys.} \textbf{\bibinfo{volume}{11}}, \bibinfo{pages}{055049}
  (\bibinfo{year}{2009}).

\bibitem[{\citenamefont{Tarbutt
  et~al.}(2009{\natexlab{a}})\citenamefont{Tarbutt, Hudson, Sauer, and
  Hinds}}]{Tarbutt(1):2009}
\bibinfo{author}{\bibfnamefont{M.~R.} \bibnamefont{Tarbutt}},
  \bibinfo{author}{\bibfnamefont{J.~J.} \bibnamefont{Hudson}},
  \bibinfo{author}{\bibfnamefont{B.~E.} \bibnamefont{Sauer}}, \bibnamefont{and}
  \bibinfo{author}{\bibfnamefont{E.~A.} \bibnamefont{Hinds}},
  \emph{\bibinfo{title}{Cold Molecules: Theory, Experiment, Applications}}
  (\bibinfo{publisher}{CRC press}, \bibinfo{year}{2009}{\natexlab{a}}),
  chap.~\bibinfo{chapter}{15}, pp. \bibinfo{pages}{555--598}.

\bibitem[{\citenamefont{Hudson et~al.}(2002)\citenamefont{Hudson, Sauer,
  Tarbutt, and Hinds}}]{Hudson:2002}
\bibinfo{author}{\bibfnamefont{J.~J.} \bibnamefont{Hudson}},
  \bibinfo{author}{\bibfnamefont{B.~E.} \bibnamefont{Sauer}},
  \bibinfo{author}{\bibfnamefont{M.~R.} \bibnamefont{Tarbutt}},
  \bibnamefont{and} \bibinfo{author}{\bibfnamefont{E.~A.} \bibnamefont{Hinds}},
  \bibinfo{journal}{Phys. Rev. Lett.} \textbf{\bibinfo{volume}{89}},
  \bibinfo{pages}{023003} (\bibinfo{year}{2002}).

\bibitem[{\citenamefont{Tarbutt
  et~al.}(2009{\natexlab{b}})\citenamefont{Tarbutt, Hudson, Sauer, and
  Hinds}}]{Tarbutt(2):2009}
\bibinfo{author}{\bibfnamefont{M.~R.} \bibnamefont{Tarbutt}},
  \bibinfo{author}{\bibfnamefont{J.~J.} \bibnamefont{Hudson}},
  \bibinfo{author}{\bibfnamefont{B.~E.} \bibnamefont{Sauer}}, \bibnamefont{and}
  \bibinfo{author}{\bibfnamefont{E.~A.} \bibnamefont{Hinds}},
  \bibinfo{journal}{Faraday Discuss.} \textbf{\bibinfo{volume}{142}},
  \bibinfo{pages}{370} (\bibinfo{year}{2009}{\natexlab{b}}).

\bibitem[{\citenamefont{Vutha et~al.}(2010)\citenamefont{Vutha, Campbell,
  Gurevich, Hutzler, Parsons, Patterson, Petrik, Spaun, Doyle, Gabrielse
  et~al.}}]{Vutha:2010}
\bibinfo{author}{\bibfnamefont{A.~C.} \bibnamefont{Vutha}},
  \bibinfo{author}{\bibfnamefont{W.~C.} \bibnamefont{Campbell}},
  \bibinfo{author}{\bibfnamefont{Y.~V.} \bibnamefont{Gurevich}},
  \bibinfo{author}{\bibfnamefont{N.~R.} \bibnamefont{Hutzler}},
  \bibinfo{author}{\bibfnamefont{M.}~\bibnamefont{Parsons}},
  \bibinfo{author}{\bibfnamefont{D.}~\bibnamefont{Patterson}},
  \bibinfo{author}{\bibfnamefont{E.}~\bibnamefont{Petrik}},
  \bibinfo{author}{\bibfnamefont{B.}~\bibnamefont{Spaun}},
  \bibinfo{author}{\bibfnamefont{J.~M.} \bibnamefont{Doyle}},
  \bibinfo{author}{\bibfnamefont{G.}~\bibnamefont{Gabrielse}},
  \bibnamefont{et~al.}, \bibinfo{journal}{J. Phys. B}
  \textbf{\bibinfo{volume}{43}}, \bibinfo{pages}{074007}
  (\bibinfo{year}{2010}).

\bibitem[{\citenamefont{Hudson et~al.}(2006)\citenamefont{Hudson, Lewandowski,
  Sawyer, and Ye}}]{Hudson:2006}
\bibinfo{author}{\bibfnamefont{E.~R.} \bibnamefont{Hudson}},
  \bibinfo{author}{\bibfnamefont{H.~J.} \bibnamefont{Lewandowski}},
  \bibinfo{author}{\bibfnamefont{B.~C.} \bibnamefont{Sawyer}},
  \bibnamefont{and} \bibinfo{author}{\bibfnamefont{J.}~\bibnamefont{Ye}},
  \bibinfo{journal}{Phys. Rev. Lett.} \textbf{\bibinfo{volume}{96}},
  \bibinfo{pages}{143004} (\bibinfo{year}{2006}).

\bibitem[{\citenamefont{van Veldhoven et~al.}(2004)\citenamefont{van Veldhoven,
  K\"upper, Bethlem, Sartakov, van Roij, and Meijer}}]{vanVeldhoven:2004}
\bibinfo{author}{\bibfnamefont{J.}~\bibnamefont{van Veldhoven}},
  \bibinfo{author}{\bibfnamefont{J.}~\bibnamefont{K\"upper}},
  \bibinfo{author}{\bibfnamefont{H.~L.} \bibnamefont{Bethlem}},
  \bibinfo{author}{\bibfnamefont{B.}~\bibnamefont{Sartakov}},
  \bibinfo{author}{\bibfnamefont{A.~J.~A.} \bibnamefont{van Roij}},
  \bibnamefont{and} \bibinfo{author}{\bibfnamefont{G.}~\bibnamefont{Meijer}},
  \bibinfo{journal}{Eur. Phys. J. D} \textbf{\bibinfo{volume}{31}},
  \bibinfo{pages}{337} (\bibinfo{year}{2004}).

\bibitem[{\citenamefont{Schiller and Korobov}(2005)}]{Schiller:2005}
\bibinfo{author}{\bibfnamefont{S.}~\bibnamefont{Schiller}} \bibnamefont{and}
  \bibinfo{author}{\bibfnamefont{V.}~\bibnamefont{Korobov}},
  \bibinfo{journal}{Phys. Rev. A} \textbf{\bibinfo{volume}{71}},
  \bibinfo{pages}{032505} (\bibinfo{year}{2005}).

\bibitem[{\citenamefont{Bethlem et~al.}(2008)\citenamefont{Bethlem, Kajita,
  Sartakov, Meijer, and Ubachs}}]{Bethlem:2008}
\bibinfo{author}{\bibfnamefont{H.~L.} \bibnamefont{Bethlem}},
  \bibinfo{author}{\bibfnamefont{M.}~\bibnamefont{Kajita}},
  \bibinfo{author}{\bibfnamefont{B.}~\bibnamefont{Sartakov}},
  \bibinfo{author}{\bibfnamefont{G.}~\bibnamefont{Meijer}}, \bibnamefont{and}
  \bibinfo{author}{\bibfnamefont{W.}~\bibnamefont{Ubachs}},
  \bibinfo{journal}{Eur. Phys. J. Special Topics}
  \textbf{\bibinfo{volume}{163}}, \bibinfo{pages}{55} (\bibinfo{year}{2008}).

\bibitem[{\citenamefont{Bethlem and Ubachs}(2009)}]{Bethlem:2009}
\bibinfo{author}{\bibfnamefont{H.~L.} \bibnamefont{Bethlem}} \bibnamefont{and}
  \bibinfo{author}{\bibfnamefont{W.}~\bibnamefont{Ubachs}},
  \bibinfo{journal}{Faraday Disscus.} \textbf{\bibinfo{volume}{142}},
  \bibinfo{pages}{25} (\bibinfo{year}{2009}).

\bibitem[{\citenamefont{M\"uller et~al.}(2004)\citenamefont{M\"uller, Herrmann,
  Saenz, Peters, and L\"ammerzahl}}]{Muller:2004}
\bibinfo{author}{\bibfnamefont{H.}~\bibnamefont{M\"uller}},
  \bibinfo{author}{\bibfnamefont{S.}~\bibnamefont{Herrmann}},
  \bibinfo{author}{\bibfnamefont{A.}~\bibnamefont{Saenz}},
  \bibinfo{author}{\bibfnamefont{A.}~\bibnamefont{Peters}}, \bibnamefont{and}
  \bibinfo{author}{\bibfnamefont{C.}~\bibnamefont{L\"ammerzahl}},
  \bibinfo{journal}{Phys. Rev. D} \textbf{\bibinfo{volume}{70}},
  \bibinfo{pages}{076004} (\bibinfo{year}{2004}).

\bibitem[{\citenamefont{DeMille et~al.}(2008)\citenamefont{DeMille, Cahn,
  Murphree, Rahmlow, and Kozlov}}]{DeMille:2008}
\bibinfo{author}{\bibfnamefont{D.}~\bibnamefont{DeMille}},
  \bibinfo{author}{\bibfnamefont{S.~B.} \bibnamefont{Cahn}},
  \bibinfo{author}{\bibfnamefont{D.}~\bibnamefont{Murphree}},
  \bibinfo{author}{\bibfnamefont{D.~A.} \bibnamefont{Rahmlow}},
  \bibnamefont{and} \bibinfo{author}{\bibfnamefont{M.~G.}
  \bibnamefont{Kozlov}}, \bibinfo{journal}{Phys. Rev. Lett.}
  \textbf{\bibinfo{volume}{100}}, \bibinfo{pages}{023003}
  (\bibinfo{year}{2008}).

\bibitem[{\citenamefont{{Darqui{\'e}} et~al.}(2010)\citenamefont{{Darqui{\'e}},
  {Stoeffler}, {Shelkovnikov}, {Daussy}, {Amy-Klein}, {Chardonnet}, {Zrig},
  {Guy}, {Crassous}, {Soulard} et~al.}}]{Darquie:2010}
\bibinfo{author}{\bibfnamefont{B.}~\bibnamefont{{Darqui{\'e}}}},
  \bibinfo{author}{\bibfnamefont{C.}~\bibnamefont{{Stoeffler}}},
  \bibinfo{author}{\bibfnamefont{A.}~\bibnamefont{{Shelkovnikov}}},
  \bibinfo{author}{\bibfnamefont{C.}~\bibnamefont{{Daussy}}},
  \bibinfo{author}{\bibfnamefont{A.}~\bibnamefont{{Amy-Klein}}},
  \bibinfo{author}{\bibfnamefont{C.}~\bibnamefont{{Chardonnet}}},
  \bibinfo{author}{\bibfnamefont{S.}~\bibnamefont{{Zrig}}},
  \bibinfo{author}{\bibfnamefont{L.}~\bibnamefont{{Guy}}},
  \bibinfo{author}{\bibfnamefont{J.}~\bibnamefont{{Crassous}}},
  \bibinfo{author}{\bibfnamefont{P.}~\bibnamefont{{Soulard}}},
  \bibnamefont{et~al.}, \bibinfo{journal}{ArXiv e-prints}
  (\bibinfo{year}{2010}), \eprint{1007.3352}.

\bibitem[{\citenamefont{Campbell and Doyle}(2009)}]{Campbell:2009}
\bibinfo{author}{\bibfnamefont{W.~C.} \bibnamefont{Campbell}} \bibnamefont{and}
  \bibinfo{author}{\bibfnamefont{J.~M.} \bibnamefont{Doyle}},
  \emph{\bibinfo{title}{Cold Molecules: Theory, Experiment, Applications}}
  (\bibinfo{publisher}{CRC press}, \bibinfo{year}{2009}),
  chap.~\bibinfo{chapter}{15}, pp. \bibinfo{pages}{473--508}.

\bibitem[{\citenamefont{Patterson et~al.}(2009)\citenamefont{Patterson,
  Rasmussen, and Doyle}}]{Patterson:2009}
\bibinfo{author}{\bibfnamefont{D.}~\bibnamefont{Patterson}},
  \bibinfo{author}{\bibfnamefont{J.}~\bibnamefont{Rasmussen}},
  \bibnamefont{and} \bibinfo{author}{\bibfnamefont{J.~M.} \bibnamefont{Doyle}},
  \bibinfo{journal}{New J. Phys.} \textbf{\bibinfo{volume}{11}},
  \bibinfo{pages}{055018} (\bibinfo{year}{2009}).

\bibitem[{\citenamefont{Weinstein
  et~al.}(1998{\natexlab{a}})\citenamefont{Weinstein, deCarvalho, Amar, Boca,
  Odom, Friedrich, and Doyle}}]{Weinstein(1):1998}
\bibinfo{author}{\bibfnamefont{J.~D.} \bibnamefont{Weinstein}},
  \bibinfo{author}{\bibfnamefont{R.}~\bibnamefont{deCarvalho}},
  \bibinfo{author}{\bibfnamefont{K.}~\bibnamefont{Amar}},
  \bibinfo{author}{\bibfnamefont{A.}~\bibnamefont{Boca}},
  \bibinfo{author}{\bibfnamefont{B.~C.} \bibnamefont{Odom}},
  \bibinfo{author}{\bibfnamefont{B.}~\bibnamefont{Friedrich}},
  \bibnamefont{and} \bibinfo{author}{\bibfnamefont{J.~M.} \bibnamefont{Doyle}},
  \bibinfo{journal}{J. Chem. Phys.} \textbf{\bibinfo{volume}{109}},
  \bibinfo{pages}{2656} (\bibinfo{year}{1998}{\natexlab{a}}).

\bibitem[{\citenamefont{Messer and De~Lucia}(1984)}]{Messer:1984}
\bibinfo{author}{\bibfnamefont{J.~K.} \bibnamefont{Messer}} \bibnamefont{and}
  \bibinfo{author}{\bibfnamefont{F.~C.} \bibnamefont{De~Lucia}},
  \bibinfo{journal}{Phys. Rev. Lett.} \textbf{\bibinfo{volume}{53}},
  \bibinfo{pages}{2555} (\bibinfo{year}{1984}).

\bibitem[{\citenamefont{Egorov et~al.}(2004)\citenamefont{Egorov, Campbell,
  Friedrich, Maxwell, Tsikata, van Buuren, and Doyle}}]{Egorov:2004}
\bibinfo{author}{\bibfnamefont{D.}~\bibnamefont{Egorov}},
  \bibinfo{author}{\bibfnamefont{W.~C.} \bibnamefont{Campbell}},
  \bibinfo{author}{\bibfnamefont{B.}~\bibnamefont{Friedrich}},
  \bibinfo{author}{\bibfnamefont{S.~E.} \bibnamefont{Maxwell}},
  \bibinfo{author}{\bibfnamefont{E.}~\bibnamefont{Tsikata}},
  \bibinfo{author}{\bibfnamefont{L.~D.} \bibnamefont{van Buuren}},
  \bibnamefont{and} \bibinfo{author}{\bibfnamefont{J.~M.} \bibnamefont{Doyle}},
  \bibinfo{journal}{Eur. Phys. J. D} \textbf{\bibinfo{volume}{31}},
  \bibinfo{pages}{307} (\bibinfo{year}{2004}).

\bibitem[{\citenamefont{Willey et~al.}(1988{\natexlab{a}})\citenamefont{Willey,
  Crownover, Bittner, and De~Lucia}}]{Willey(1):1988}
\bibinfo{author}{\bibfnamefont{D.~R.} \bibnamefont{Willey}},
  \bibinfo{author}{\bibfnamefont{R.~L.} \bibnamefont{Crownover}},
  \bibinfo{author}{\bibfnamefont{D.~N.} \bibnamefont{Bittner}},
  \bibnamefont{and} \bibinfo{author}{\bibfnamefont{F.~C.}
  \bibnamefont{De~Lucia}}, \bibinfo{journal}{J. Chem. Phys.}
  \textbf{\bibinfo{volume}{89}}, \bibinfo{pages}{1923}
  (\bibinfo{year}{1988}{\natexlab{a}}).

\bibitem[{\citenamefont{Willey et~al.}(1988{\natexlab{b}})\citenamefont{Willey,
  Crownover, Bittner, and De~Lucia}}]{Willey(2):1988}
\bibinfo{author}{\bibfnamefont{D.~R.} \bibnamefont{Willey}},
  \bibinfo{author}{\bibfnamefont{R.~L.} \bibnamefont{Crownover}},
  \bibinfo{author}{\bibfnamefont{D.~N.} \bibnamefont{Bittner}},
  \bibnamefont{and} \bibinfo{author}{\bibfnamefont{F.~C.}
  \bibnamefont{De~Lucia}}, \bibinfo{journal}{J. Chem. Phys.}
  \textbf{\bibinfo{volume}{89}}, \bibinfo{pages}{6147}
  (\bibinfo{year}{1988}{\natexlab{b}}).

\bibitem[{\citenamefont{Egorov et~al.}(2001)\citenamefont{Egorov, Weinstein,
  Patterson, Friedrich, and Doyle}}]{Egorov:2001}
\bibinfo{author}{\bibfnamefont{D.}~\bibnamefont{Egorov}},
  \bibinfo{author}{\bibfnamefont{J.~D.} \bibnamefont{Weinstein}},
  \bibinfo{author}{\bibfnamefont{D.}~\bibnamefont{Patterson}},
  \bibinfo{author}{\bibfnamefont{B.}~\bibnamefont{Friedrich}},
  \bibnamefont{and} \bibinfo{author}{\bibfnamefont{J.~M.} \bibnamefont{Doyle}},
  \bibinfo{journal}{Phys. Rev. A} \textbf{\bibinfo{volume}{63}},
  \bibinfo{pages}{030501(R)} (\bibinfo{year}{2001}).

\bibitem[{\citenamefont{Skoff et~al.}(2009)\citenamefont{Skoff, Hendricks,
  Sinclair, Tarbutt, Hudson, Segal, Sauer, and Hinds}}]{Skoff:2009}
\bibinfo{author}{\bibfnamefont{S.~M.} \bibnamefont{Skoff}},
  \bibinfo{author}{\bibfnamefont{R.~J.} \bibnamefont{Hendricks}},
  \bibinfo{author}{\bibfnamefont{C.~D.~J.} \bibnamefont{Sinclair}},
  \bibinfo{author}{\bibfnamefont{M.~R.} \bibnamefont{Tarbutt}},
  \bibinfo{author}{\bibfnamefont{J.~J.} \bibnamefont{Hudson}},
  \bibinfo{author}{\bibfnamefont{D.~M.} \bibnamefont{Segal}},
  \bibinfo{author}{\bibfnamefont{B.~E.} \bibnamefont{Sauer}}, \bibnamefont{and}
  \bibinfo{author}{\bibfnamefont{E.~A.} \bibnamefont{Hinds}},
  \bibinfo{journal}{New J. Phys.} \textbf{\bibinfo{volume}{11}},
  \bibinfo{pages}{123026} (\bibinfo{year}{2009}).

\bibitem[{\citenamefont{Lu et~al.}(2008)\citenamefont{Lu, Hardman, Weinstein,
  and Zygleman}}]{Lu:2008}
\bibinfo{author}{\bibfnamefont{M.-J.} \bibnamefont{Lu}},
  \bibinfo{author}{\bibfnamefont{K.~S.} \bibnamefont{Hardman}},
  \bibinfo{author}{\bibfnamefont{J.~D.} \bibnamefont{Weinstein}},
  \bibnamefont{and} \bibinfo{author}{\bibfnamefont{B.}~\bibnamefont{Zygleman}},
  \bibinfo{journal}{Phys. Rev. A} \textbf{\bibinfo{volume}{77}},
  \bibinfo{pages}{060701(R)} (\bibinfo{year}{2008}).

\bibitem[{\citenamefont{Lu and Weinstein}(2009)}]{Lu:2009}
\bibinfo{author}{\bibfnamefont{M.-J.} \bibnamefont{Lu}} \bibnamefont{and}
  \bibinfo{author}{\bibfnamefont{J.~D.} \bibnamefont{Weinstein}},
  \bibinfo{journal}{New J. Phys.} \textbf{\bibinfo{volume}{11}},
  \bibinfo{pages}{055015} (\bibinfo{year}{2009}).

\bibitem[{\citenamefont{Doyle et~al.}(1995)\citenamefont{Doyle, Friedrich, Kim,
  and Patterson}}]{Doyle:1995}
\bibinfo{author}{\bibfnamefont{J.~M.} \bibnamefont{Doyle}},
  \bibinfo{author}{\bibfnamefont{B.}~\bibnamefont{Friedrich}},
  \bibinfo{author}{\bibfnamefont{J.}~\bibnamefont{Kim}}, \bibnamefont{and}
  \bibinfo{author}{\bibfnamefont{D.}~\bibnamefont{Patterson}},
  \bibinfo{journal}{Phys. Rev. A} \textbf{\bibinfo{volume}{52}},
  \bibinfo{pages}{R2515} (\bibinfo{year}{1995}).

\bibitem[{\citenamefont{Weinstein
  et~al.}(1998{\natexlab{b}})\citenamefont{Weinstein, deCarvalho, Guillet,
  Friedrich, and Doyle}}]{Weinstein(2):1998}
\bibinfo{author}{\bibfnamefont{J.~D.} \bibnamefont{Weinstein}},
  \bibinfo{author}{\bibfnamefont{R.}~\bibnamefont{deCarvalho}},
  \bibinfo{author}{\bibfnamefont{T.}~\bibnamefont{Guillet}},
  \bibinfo{author}{\bibfnamefont{B.}~\bibnamefont{Friedrich}},
  \bibnamefont{and} \bibinfo{author}{\bibfnamefont{J.~M.} \bibnamefont{Doyle}},
  \bibinfo{journal}{Nature} \textbf{\bibinfo{volume}{395}},
  \bibinfo{pages}{148} (\bibinfo{year}{1998}{\natexlab{b}}).

\bibitem[{\citenamefont{Stoll et~al.}(2008)\citenamefont{Stoll, Bakker,
  Steimle, Meijer, and Peters}}]{Stoll:2008}
\bibinfo{author}{\bibfnamefont{M.}~\bibnamefont{Stoll}},
  \bibinfo{author}{\bibfnamefont{J.~M.} \bibnamefont{Bakker}},
  \bibinfo{author}{\bibfnamefont{T.~C.} \bibnamefont{Steimle}},
  \bibinfo{author}{\bibfnamefont{G.}~\bibnamefont{Meijer}}, \bibnamefont{and}
  \bibinfo{author}{\bibfnamefont{A.}~\bibnamefont{Peters}},
  \bibinfo{journal}{Phys. Rev. A} \textbf{\bibinfo{volume}{78}},
  \bibinfo{pages}{032707} (\bibinfo{year}{2008}).

\bibitem[{\citenamefont{Maussang et~al.}(2005)\citenamefont{Maussang, Egorov,
  Helton, Nguyen, and Doyle}}]{Maussang:2005}
\bibinfo{author}{\bibfnamefont{K.}~\bibnamefont{Maussang}},
  \bibinfo{author}{\bibfnamefont{D.}~\bibnamefont{Egorov}},
  \bibinfo{author}{\bibfnamefont{J.~S.} \bibnamefont{Helton}},
  \bibinfo{author}{\bibfnamefont{S.~V.} \bibnamefont{Nguyen}},
  \bibnamefont{and} \bibinfo{author}{\bibfnamefont{J.~M.} \bibnamefont{Doyle}},
  \bibinfo{journal}{Phys. Rev. Lett.} \textbf{\bibinfo{volume}{94}},
  \bibinfo{pages}{123002} (\bibinfo{year}{2005}).

\bibitem[{\citenamefont{Tskita et~al.}(2010)\citenamefont{Tskita, Campbell,
  Hummon, Lu, and Doyle}}]{Tsikata:2010}
\bibinfo{author}{\bibfnamefont{E.}~\bibnamefont{Tskita}},
  \bibinfo{author}{\bibfnamefont{W.~C.} \bibnamefont{Campbell}},
  \bibinfo{author}{\bibfnamefont{M.~T.} \bibnamefont{Hummon}},
  \bibinfo{author}{\bibfnamefont{H.-I.} \bibnamefont{Lu}}, \bibnamefont{and}
  \bibinfo{author}{\bibfnamefont{J.~M.} \bibnamefont{Doyle}},
  \bibinfo{journal}{New. J. Phys.} \textbf{\bibinfo{volume}{12}},
  \bibinfo{pages}{065028} (\bibinfo{year}{2010}).

\bibitem[{\citenamefont{Maxwell et~al.}(2005)\citenamefont{Maxwell, Brahms,
  deCarvalho, Glenn, Helton, Nguyen, Patterson, Petricka, DeMille, and
  Doyle}}]{Maxwell:2005}
\bibinfo{author}{\bibfnamefont{S.~E.} \bibnamefont{Maxwell}},
  \bibinfo{author}{\bibfnamefont{N.}~\bibnamefont{Brahms}},
  \bibinfo{author}{\bibfnamefont{R.}~\bibnamefont{deCarvalho}},
  \bibinfo{author}{\bibfnamefont{D.~R.} \bibnamefont{Glenn}},
  \bibinfo{author}{\bibfnamefont{J.~S.} \bibnamefont{Helton}},
  \bibinfo{author}{\bibfnamefont{S.~V.} \bibnamefont{Nguyen}},
  \bibinfo{author}{\bibfnamefont{D.}~\bibnamefont{Patterson}},
  \bibinfo{author}{\bibfnamefont{J.}~\bibnamefont{Petricka}},
  \bibinfo{author}{\bibfnamefont{D.}~\bibnamefont{DeMille}}, \bibnamefont{and}
  \bibinfo{author}{\bibfnamefont{J.~M.} \bibnamefont{Doyle}},
  \bibinfo{journal}{Phys. Rev. Lett.} \textbf{\bibinfo{volume}{95}},
  \bibinfo{pages}{173201} (\bibinfo{year}{2005}).

\bibitem[{\citenamefont{Patterson and Doyle}(2007)}]{Patterson:2007}
\bibinfo{author}{\bibfnamefont{D.}~\bibnamefont{Patterson}} \bibnamefont{and}
  \bibinfo{author}{\bibfnamefont{J.~M.} \bibnamefont{Doyle}},
  \bibinfo{journal}{J. Chem. Phys.} \textbf{\bibinfo{volume}{126}},
  \bibinfo{pages}{154307} (\bibinfo{year}{2007}).

\bibitem[{\citenamefont{van Buuren et~al.}(2009)\citenamefont{van Buuren,
  Sommer, Motsch, Pohle, Schenk, Bayerl, Pinkse, and Rempe}}]{vanBuuren:2009}
\bibinfo{author}{\bibfnamefont{L.~D.} \bibnamefont{van Buuren}},
  \bibinfo{author}{\bibfnamefont{C.}~\bibnamefont{Sommer}},
  \bibinfo{author}{\bibfnamefont{M.}~\bibnamefont{Motsch}},
  \bibinfo{author}{\bibfnamefont{S.}~\bibnamefont{Pohle}},
  \bibinfo{author}{\bibfnamefont{M.}~\bibnamefont{Schenk}},
  \bibinfo{author}{\bibfnamefont{J.}~\bibnamefont{Bayerl}},
  \bibinfo{author}{\bibfnamefont{P.~W.~H.} \bibnamefont{Pinkse}},
  \bibnamefont{and} \bibinfo{author}{\bibfnamefont{G.}~\bibnamefont{Rempe}},
  \bibinfo{journal}{Phys. Rev. Lett.} \textbf{\bibinfo{volume}{102}},
  \bibinfo{pages}{033001} (\bibinfo{year}{2009}).

\bibitem[{\citenamefont{Wall et~al.}(2009{\natexlab{a}})\citenamefont{Wall,
  Armitage, Hudson, Sauer, Dyne, Hinds, and Tarbutt}}]{Wall:2009}
\bibinfo{author}{\bibfnamefont{T.~E.} \bibnamefont{Wall}},
  \bibinfo{author}{\bibfnamefont{S.}~\bibnamefont{Armitage}},
  \bibinfo{author}{\bibfnamefont{J.~J.} \bibnamefont{Hudson}},
  \bibinfo{author}{\bibfnamefont{B.~E.} \bibnamefont{Sauer}},
  \bibinfo{author}{\bibfnamefont{J.~M.} \bibnamefont{Dyne}},
  \bibinfo{author}{\bibfnamefont{E.~A.} \bibnamefont{Hinds}}, \bibnamefont{and}
  \bibinfo{author}{\bibfnamefont{M.~R.} \bibnamefont{Tarbutt}},
  \bibinfo{journal}{Phys. Rev. A} \textbf{\bibinfo{volume}{80}},
  \bibinfo{pages}{043407} (\bibinfo{year}{2009}{\natexlab{a}}).

\bibitem[{\citenamefont{Roberts and Sydoriak}(1956)}]{Roberts:1956}
\bibinfo{author}{\bibfnamefont{R.~P.} \bibnamefont{Roberts}} \bibnamefont{and}
  \bibinfo{author}{\bibfnamefont{S.~G.} \bibnamefont{Sydoriak}},
  \bibinfo{journal}{Phys. Rev.} \textbf{\bibinfo{volume}{102}},
  \bibinfo{pages}{304} (\bibinfo{year}{1956}).

\bibitem[{\citenamefont{Chapman}(1916)}]{Chapman:1916}
\bibinfo{author}{\bibfnamefont{S.}~\bibnamefont{Chapman}},
  \bibinfo{journal}{Proc. Royal Soc. London} \textbf{\bibinfo{volume}{93}},
  \bibinfo{pages}{1} (\bibinfo{year}{1916}).

\bibitem[{\citenamefont{Hasted}(1972)}]{Hasted:1972}
\bibinfo{author}{\bibnamefont{Hasted}}, \emph{\bibinfo{title}{Physics of Atomic
  Collisions}} (\bibinfo{publisher}{Butterworths}, \bibinfo{year}{1972}),
  \bibinfo{edition}{2nd} ed.

\bibitem[{\citenamefont{Mason and Marrero}(1970)}]{Mason:1970}
\bibinfo{author}{\bibfnamefont{E.~A.} \bibnamefont{Mason}} \bibnamefont{and}
  \bibinfo{author}{\bibfnamefont{T.~R.} \bibnamefont{Marrero}},
  \bibinfo{journal}{Advances in Atomic and Molecular Physics}
  \textbf{\bibinfo{volume}{6}}, \bibinfo{pages}{155} (\bibinfo{year}{1970}).

\bibitem[{\citenamefont{Czuchaj et~al.}(1995)\citenamefont{Czuchaj, Rebentrost,
  Stoll, and Preuss}}]{Czuchaj:1995}
\bibinfo{author}{\bibfnamefont{E.}~\bibnamefont{Czuchaj}},
  \bibinfo{author}{\bibfnamefont{F.}~\bibnamefont{Rebentrost}},
  \bibinfo{author}{\bibfnamefont{H.}~\bibnamefont{Stoll}}, \bibnamefont{and}
  \bibinfo{author}{\bibfnamefont{H.}~\bibnamefont{Preuss}},
  \bibinfo{journal}{Chem. Phys.} \textbf{\bibinfo{volume}{196}},
  \bibinfo{pages}{37} (\bibinfo{year}{1995}).

\bibitem[{\citenamefont{Patil}(1991)}]{Patil:1991}
\bibinfo{author}{\bibfnamefont{S.~H.} \bibnamefont{Patil}},
  \bibinfo{journal}{J. Chem. Phys.} \textbf{\bibinfo{volume}{94}},
  \bibinfo{pages}{8089} (\bibinfo{year}{1991}).

\bibitem[{\citenamefont{Tscherbul et~al.}(2007)\citenamefont{Tscherbul,
  K{\l}os, Rajchel, and Krems}}]{Tscherbul:2007}
\bibinfo{author}{\bibfnamefont{T.~V.} \bibnamefont{Tscherbul}},
  \bibinfo{author}{\bibfnamefont{J.}~\bibnamefont{K{\l}os}},
  \bibinfo{author}{\bibfnamefont{L.}~\bibnamefont{Rajchel}}, \bibnamefont{and}
  \bibinfo{author}{\bibfnamefont{R.~V.} \bibnamefont{Krems}},
  \bibinfo{journal}{Phys. Rev. A} \textbf{\bibinfo{volume}{75}},
  \bibinfo{pages}{033416} (\bibinfo{year}{2007}).

\bibitem[{\citenamefont{Sushkov and Budker}(2008)}]{Sushkov:2008}
\bibinfo{author}{\bibfnamefont{A.~O.} \bibnamefont{Sushkov}} \bibnamefont{and}
  \bibinfo{author}{\bibfnamefont{D.}~\bibnamefont{Budker}},
  \bibinfo{journal}{Phys. Rev. A} \textbf{\bibinfo{volume}{77}},
  \bibinfo{pages}{042707} (\bibinfo{year}{2008}).

\bibitem[{\citenamefont{deCarvalho et~al.}(1999)\citenamefont{deCarvalho,
  Doyle, Friedrich, Guillet, Kim, Patterson, and Weinstein}}]{deCarvalho:1999}
\bibinfo{author}{\bibfnamefont{R.}~\bibnamefont{deCarvalho}},
  \bibinfo{author}{\bibfnamefont{J.~M.} \bibnamefont{Doyle}},
  \bibinfo{author}{\bibfnamefont{B.}~\bibnamefont{Friedrich}},
  \bibinfo{author}{\bibfnamefont{T.}~\bibnamefont{Guillet}},
  \bibinfo{author}{\bibfnamefont{J.}~\bibnamefont{Kim}},
  \bibinfo{author}{\bibfnamefont{D.}~\bibnamefont{Patterson}},
  \bibnamefont{and} \bibinfo{author}{\bibfnamefont{J.~D.}
  \bibnamefont{Weinstein}}, \bibinfo{journal}{Eur. Phys. J. D}
  \textbf{\bibinfo{volume}{7}}, \bibinfo{pages}{289} (\bibinfo{year}{1999}).

\bibitem[{\citenamefont{Wall et~al.}(2008{\natexlab{b}})\citenamefont{Wall,
  Kanem, Hudson, Sauer, Cho, Boshier, Hinds, and Tarbutt}}]{Wall:2008}
\bibinfo{author}{\bibfnamefont{T.~E.} \bibnamefont{Wall}},
  \bibinfo{author}{\bibfnamefont{J.~F.} \bibnamefont{Kanem}},
  \bibinfo{author}{\bibfnamefont{J.~J.} \bibnamefont{Hudson}},
  \bibinfo{author}{\bibfnamefont{B.~E.} \bibnamefont{Sauer}},
  \bibinfo{author}{\bibfnamefont{D.}~\bibnamefont{Cho}},
  \bibinfo{author}{\bibfnamefont{M.~G.} \bibnamefont{Boshier}},
  \bibinfo{author}{\bibfnamefont{E.~A.} \bibnamefont{Hinds}}, \bibnamefont{and}
  \bibinfo{author}{\bibfnamefont{M.~R.} \bibnamefont{Tarbutt}},
  \bibinfo{journal}{Phys. Rev. A} \textbf{\bibinfo{volume}{78}},
  \bibinfo{pages}{062509} (\bibinfo{year}{2008}{\natexlab{b}}).

\bibitem[{\citenamefont{Tarbutt et~al.}(2002)\citenamefont{Tarbutt, Hudson,
  Sauer, Hinds, Ryzhov, Ryabov, and Ezhov}}]{Tarbutt:2002}
\bibinfo{author}{\bibfnamefont{M.~R.} \bibnamefont{Tarbutt}},
  \bibinfo{author}{\bibfnamefont{J.~J.} \bibnamefont{Hudson}},
  \bibinfo{author}{\bibfnamefont{B.~E.} \bibnamefont{Sauer}},
  \bibinfo{author}{\bibfnamefont{E.~A.} \bibnamefont{Hinds}},
  \bibinfo{author}{\bibfnamefont{V.~A.} \bibnamefont{Ryzhov}},
  \bibinfo{author}{\bibfnamefont{V.~L.} \bibnamefont{Ryabov}},
  \bibnamefont{and} \bibinfo{author}{\bibfnamefont{V.~F.} \bibnamefont{Ezhov}},
  \bibinfo{journal}{J. Phys. B} \textbf{\bibinfo{volume}{35}},
  \bibinfo{pages}{5013} (\bibinfo{year}{2002}).

\bibitem[{\citenamefont{Pelegrini et~al.}(2005)\citenamefont{Pelegrini,
  Vivacqua, Roberto-Neto, Ornellas, and Machado}}]{Pelegrini:2005}
\bibinfo{author}{\bibfnamefont{M.}~\bibnamefont{Pelegrini}},
  \bibinfo{author}{\bibfnamefont{C.~S.} \bibnamefont{Vivacqua}},
  \bibinfo{author}{\bibfnamefont{O.}~\bibnamefont{Roberto-Neto}},
  \bibinfo{author}{\bibfnamefont{F.~R.} \bibnamefont{Ornellas}},
  \bibnamefont{and} \bibinfo{author}{\bibfnamefont{F.~B.~C.}
  \bibnamefont{Machado}}, \bibinfo{journal}{Braz. J. Phys.}
  \textbf{\bibinfo{volume}{35}}, \bibinfo{pages}{950} (\bibinfo{year}{2005}).

\bibitem[{\citenamefont{Smith and H\"ansch}(1971)}]{Smith:1971}
\bibinfo{author}{\bibfnamefont{P.~W.} \bibnamefont{Smith}} \bibnamefont{and}
  \bibinfo{author}{\bibfnamefont{T.}~\bibnamefont{H\"ansch}},
  \bibinfo{journal}{Phys. Rev. Lett.} \textbf{\bibinfo{volume}{26}},
  \bibinfo{pages}{740} (\bibinfo{year}{1971}).

\end{thebibliography}
\end{document}